\documentclass[aps,prb,twocolumn,10pt,amssymb,amsmath,superscriptaddress,longbibliography]{revtex4-2} 
\usepackage{amsfonts,amsmath,amssymb,graphicx,stmaryrd,chemformula,gensymb}
\usepackage{times}
\usepackage{xcolor}
\usepackage{grffile}
\usepackage{braket}
\usepackage{bm}
\usepackage{subdepth} % subscripts on same level even if there are some superscripts
\usepackage[version=4]{mhchem}
\usepackage{booktabs}
\usepackage{siunitx}
\usepackage{tabularx}
\usepackage{svg}
\usepackage{url}
\definecolor{darkblue}{rgb}{0, 0, 0.8}
\usepackage[colorlinks=true, breaklinks=true, linkcolor=darkblue, citecolor=darkblue, urlcolor=darkblue]{hyperref}

\renewcommand{\eqref}[1]{Eq.~(\ref{#1})}

\newcommand{\muB}{\mu_{\rm{B}}}

\def\nnb{\langle i,j \rangle}

\graphicspath{{./figures/}}

\begin{document}

\title{Field-Induced Magnon Decay, Magnon Shadows, and Roton-like Excitations in the Honeycomb Antiferromagnet YbBr$_3$}

\author{J. A. Hernández} 
\affiliation{PSI Center for Neutron and Muon Sciences, CH-5232 Villigen-PSI, Switzerland}

\author{A. A. Eberharter} 
\affiliation{PSI Center for Scientific Computing, Theory and Data, CH-5232 Villigen-PSI, Switzerland}

\author{M. Schuler}
\affiliation{Institut f\"ur Theoretische Physik, Universit\"at Innsbruck, A-6020 Innsbruck, Austria}

\author{J. Lass}
\affiliation{PSI Center for Neutron and Muon Sciences, CH-5232 Villigen-PSI, Switzerland}

\author{D. G. Mazzone}
\affiliation{PSI Center for Neutron and Muon Sciences, CH-5232 Villigen-PSI, Switzerland}

\author{R. Sibille} 
\affiliation{PSI Center for Neutron and Muon Sciences, CH-5232 Villigen-PSI, Switzerland}

\author{S. Raymond}
\affiliation{Université Grenoble Alpes, CEA, IRIG, MEM, MDN, 38000 Grenoble, France}

\author{K. W. Krämer}
\affiliation{Department of Chemistry, Biochemistry and Pharmaceutical Sciences, University 
of Bern, Freiestrasse 3, CH-3012 Bern, Switzerland.}

\author{B. Normand}
\affiliation{PSI Center for Scientific Computing, Theory and Data, CH-5232 Villigen-PSI, Switzerland}
\affiliation{Institute of Physics, Ecole Polytechnique Fédérale de Lausanne (EPFL), CH-1015 Lausanne, Switzerland}

\author{B. Roessli}
\affiliation{PSI Center for Neutron and Muon Sciences, CH-5232 Villigen-PSI, Switzerland}

\author{A. M. Läuchli}
\affiliation{PSI Center for Scientific Computing, Theory and Data, CH-5232 Villigen-PSI, Switzerland}
\affiliation{Institute of Physics, Ecole Polytechnique Fédérale de Lausanne (EPFL), CH-1015 Lausanne, Switzerland}

\author{M. Kenzelmann} 
\affiliation{PSI Center for Neutron and Muon Sciences, CH-5232 Villigen-PSI, Switzerland}

\begin{abstract}
Although the search for quantum many-body phenomena in magnetic materials has a strong focus on highly frustrated systems, even unfrustrated quantum magnets show a multitude of unconventional phenomena in their spin excitation spectra. \ce{YbBr3} is an excellent realization of the $S = 1/2$ antiferromagnetic Heisenberg model on the honeycomb lattice, and we have performed detailed spectroscopic experiments with both unpolarized and polarized neutrons at all applied magnetic fields up to saturation. We observe extensive excitation continua, which cause strong renormalization and the decay of single magnons at higher fields, while coherent features include field-induced ``shadows'' of the single magnons and the spectacular emergence of a roton-like excitation. To guide and interpret our experiments, we performed systematic calculations by the method of cylinder matrix-product states that provide quantitative agreement with the neutron scattering data and a qualitative benchmark for the spectral signatures of strong quantum fluctuations even in the absence of magnetic frustration.
\end{abstract}

\maketitle

\emph{Introduction.}---Quantum magnetic models and materials provide an excellent proving ground for highly entangled quantum many-body phenomena. Much of this research has focused on systems with low dimensionality and high frustration, such that quantum fluctuation effects cause the complete destabilization of magnetic order, and on fingerprints of the resulting ``quantum spin liquid'' (QSL) \cite{Savary2016,Broholm2020}. Much more common, however, is ``renormalized classical'' behavior, in which ordered magnetic properties including the sublattice magnetization, spin stiffness, and spin-wave velocity are finite, but strongly renormalized by quantum fluctuations. Nevertheless, some well documented quantum phenomena in unfrustrated systems question whether a renormalization description can be sufficient, or if some hallmarks of the QSL may appear at finite energies. Key examples for the paradigm square-lattice Heisenberg antiferromagnet (SLHAF) are the theoretical discussion of magnon decay at high fields \cite{zc1999,Luescher2009,mzc2010,fmzc2012,zc2013} and the experimental observation of the $(\pi,0)$ scattering anomaly \cite{Headings2010,Plumb2014,DallaPiazza2014}, which sharpened the question of whether the spectral function reflects many multimagnon bound and scattering states \cite{Powalski2016,Powalski2018} or a magnon fractionalization \cite{AA1988,Hsu1990,Ho2001,DallaPiazza2014,Shao2017,Yu2018,Zhang2022}. 

\begin{figure}[t]
\includegraphics[width=\linewidth]{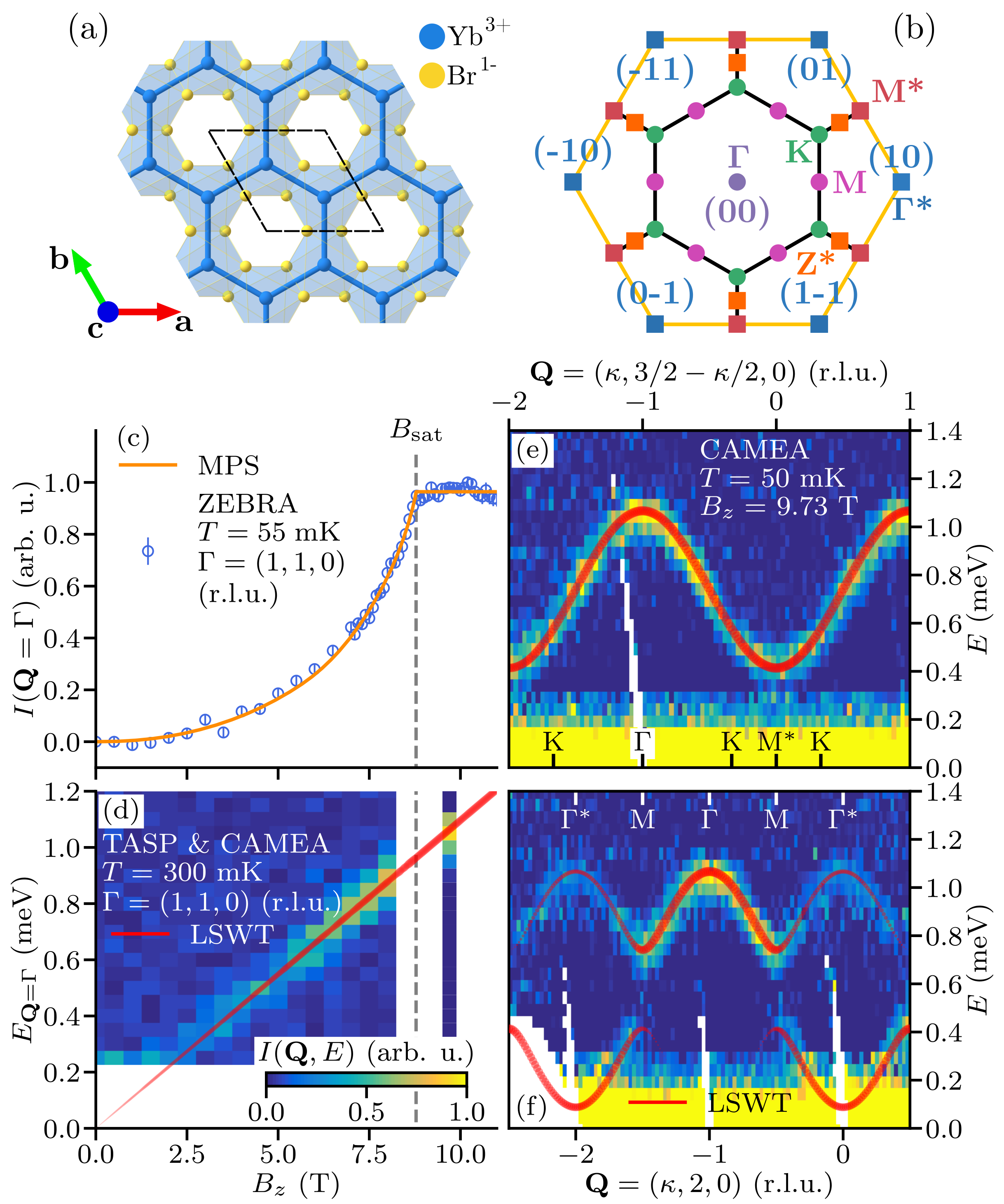}
\caption{\label{fig:hp} {\bf Crystal structure and interaction parameters.} (a) Crystal structure of a single YbBr$_3$ layer. Edge-sharing \ce{Br}$^{-}$ octahedra surround the magnetic \ce{Yb}$^{3+}$ ions. (b) Reciprocal space for the honeycomb lattice, showing the Brillouin zone (BZ, black hexagon) and irreducible BZ (orange hexagon). (c) Complete magnetization response deduced from neutron diffraction measurements of the ordered moment at $T = 55$ mK. The orange line is the result calculated by MPS.
(d) Measurements at the $\Gamma$ point showing the Larmor-mode frequency, $\hbar \omega_{\rm L} = g_{zz} \muB B_z$. The red line is the result calculated by linear spin-wave theory (LSWT). The vertical dashed line marks the deduced saturation field, $B_{\rm sat} = 8.78(2)$ T. (e-f) INS spectra measured in the fully polarized phase, at $B_z = 9.73$ T, along the two high-symmetry directions in the BZ. Red lines indicate the optimal fit by LSWT used to extract the interaction parameters. The widths of the LSWT lines are proportional to the calculated intensities.}
\end{figure}

One of the ``most quantum'' unfrustrated systems on a regular two-dimensional (2D) geometry is the $S = 1/2$ honeycomb lattice, with coordination number $z = 3$. Although the honeycomb Heisenberg model was studied intensively for its ground-state phase diagram in the presence of frustrating next- ($J_2$) and third-neighbor ($J_3$) coupling \cite{Fouet2001,Mulder2010,Oitmaa2011,Farnell2011,Albuquerque2011,Ganesh2013,Zhu2013,Gong2013,Ghorbani2016}, and also for the topological magnon excitations of the nearest-neighbor ($J_1$) model \cite{Pershoguba2018,McClarty2022,Nikitin2022}, studies of the excitation spectrum have only recently been extended from the semiclassical level \cite{mc2016} to discuss many-body quantum phenomena \cite{Ferrari2020,Gu2022}. Layered van der Waals materials offer the possibility of measuring the magnetic properties of 2D systems, and the $AX_3$ family of lanthanide ($A$) trihalides ($X = \rm{Cl}, \rm{Br}, \rm{I}$) offers two ideal properties, namely the precise honeycomb geometry shown in Fig.~\ref{fig:hp}(a) and, for the \ce{Yb}-based systems, effective spin $1/2$ and saturation fields ($B_{\rm sat}$) within range of experiment. Early measurements on \ce{YbBr3} found a complex zero-field spectrum \cite{Wessler2020}, while studies of \ce{YbCl3} discussed one-magnon physics \cite{Sala2021} and sharp features in the scattering continua ascribed to two noninteracting magnons \cite{Sala2023}, but achieved only a partial account of the observed spectra.

In this Letter we report our comprehensive investigation of the honeycomb Heisenberg antiferromagnet (HHAF) \ce{YbBr3} by inelastic neutron scattering (INS) experiments and cylinder matrix-product-states (MPS) calculations of the spectral function. We find excellent quantitative agreement between the measured INS data and an MPS treatment of the nearest-neighbor HHAF for all applied magnetic fields up to saturation. At finite fields we find four remarkable phenomena: the field-induced restoration of a magnon-like excitation from the continuum scattering at the K point of the Brillouin zone (BZ); the almost complete field-induced destruction of magnon-like excitations close to the critical field, caused by spectral-weight transfer into broad continua; an additional excitation at low fields that ``shadows'' the one-magnon modes; a nonmonotonic one-magnon dispersion reminiscent of the roton in superfluid helium. Having benchmarked the complete spectral function, we discuss the candidate theoretical descriptions accounting for the coexistence of the magnon-like excitations and complex continua we observe.

\emph{Material, experiments, and methods.}---For our study we used the high quality single-crystal samples of \ce{YbBr3} [Fig.~\ref{fig:hp}(a)] measured at zero field in Ref.~\cite{Wessler2020}, whose growth and structurally 2D character we summarize in Sec.~S1 of the Supplemental Material (SM) \cite{sm}. We performed unpolarized INS experiments to measure the magnetic excitations at all fields on the cold-neutron triple-axis spectrometer (TAS) TASP and the multiplexing spectrometer CAMEA \cite{Lass2023} at the Swiss Spallation Neutron Source, SINQ, at the Paul Scherrer Institut (PSI), as well as polarized INS on the TAS IN12 at the Institut Laue Langevin (ILL) \cite{Schmalzl2016}; these measurements and the accompanying data analysis are detailed in Sec.~S2 of the SM \cite{sm}. The low energy scale of the magnetic interactions in \ce{YbBr3} demands precision measurements at sub-Kelvin temperatures in superconducting cryomagnets. However, it also allows experimental access to the saturation field, which we determined to high accuracy by measuring the intensity of the $\mathbf{Q} = \Gamma$ reflection as a function of magnetic field applied along the crystalline $c$ axis on the thermal neutron diffractometer ZEBRA (PSI), as shown in Fig.~\ref{fig:hp}(c), obtaining the result $B_{\rm{sat}} = 8.78(2)$ T.

To interpret our INS spectra, we performed extensive calculations by the MPS method, which provides near-unbiased spectral functions for 2D systems wrapped on a cylinder \cite{Gohlke2017,Verresen2018,Xie2023}. Section S4 of the SM \cite{sm} explains the application and benchmarking of cylinder MPS for the HHAF, extraction of the dynamical structure factor from MPS correlation functions, and resolution convolution using \textsc{Takin} \cite{Weber2016} by which we achieved a fully quantitative comparison of INS and MPS.

\emph{Fully polarized phase.}---All quantum fluctuations are suppressed in the fully polarized phase above $B_{\rm{sat}}$. Any spin-flip excitation is then described exactly by linear spin-wave theory (LSWT), whose implementation we summarize in Sec.~S3 of the SM \cite{sm}. This makes it possible to determine very precisely the microscopic interaction parameters that govern the low-energy physics of \ce{YbBr3} at all fields (taking magnetostriction effects in octahedrally coordinated Yb$^{3+}$ materials to be negligible below 10 T \cite{Pocs2021}). We remark that any Heisenberg magnet has an excitation at $\mathbf{Q} = \Gamma$ whose energy ($\hbar \omega_{\rm L}$) rises linearly with the field \cite{oshikawa1997}. As Fig.~\ref{fig:hp}(d) shows, this Larmor mode, or precession mode, is clearly visible and very accurately linear in \ce{YbBr3}, providing no evidence for magnetostriction or non-Heisenberg terms in the spin Hamiltonian. 

\begin{figure*}[t]
\includegraphics[width=\linewidth]{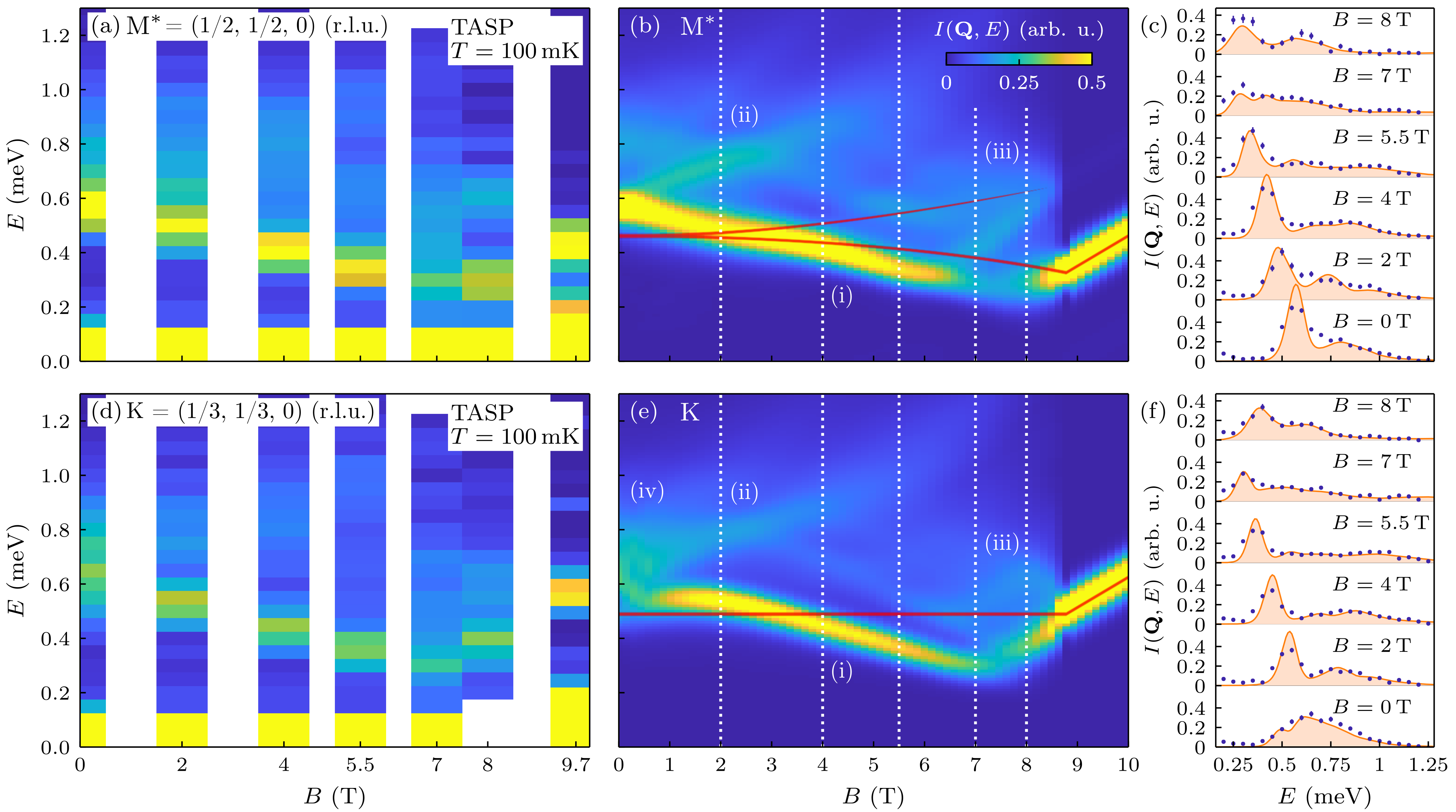}
\caption{\label{fig:Hdep}{\bf Field-dependence of magnetic excitations.} (a) INS measurements and (b) MPS calculations of the field-induced spectral function at the M$^*$ point. (c) Quantitative comparison of INS and MPS at M$^*$ for all fields applied in experiment. (d) INS measurements and (e) MPS calculations at the K point. (f) Comparison of INS and MPS at K. Red lines in panels (b) and (e) show the magnon branches of LSWT. The labels (i), (ii), (iii), and (iv) in panels (b) and (e) refer to the specific excitation features discussed in the text.}
\end{figure*}

Working above $B_{\rm sat}$, in a field $B_{z} = 9.73$ T applied perpendicular to the honeycomb plane, we measured the magnon dispersion at $T = 50$ mK along the two high-symmetry directions of the BZ as shown in Figs.~\ref{fig:hp}(e-f). Neglecting single-ion anisotropies for an $S_{\rm{eff}} = 1/2$ system, we fit these data to a $J_1$-$J_2$-$J_3$-$\Delta$ model with Heisenberg interactions between the first three near-neighbor sites and an additive XXZ anisotropy on the $J_1$ bond \big[making the full $z$-component term $\sum_{\nnb} (J_1+\Delta) \hat{S}_i^z \hat{S}_j^z$\big]. The optimal values of $J_2$, $J_3$, and $\Delta$ are all zero within the precision afforded by the INS data, as we show in detail in Sec.~S3 of the SM \cite{sm}. An appropriate minimal model to describe the measured dispersion is therefore the pure nearest-neighbor (n.n.) HHAF Hamiltonian, 
\begin{equation}\label{eq:hamiltonian}
    \hat{\mathcal{H}} = J_1 \sum_{\nnb} \hat{\mathbf{S}}_{i} \! \cdot \! \hat{\mathbf{S}}_{j} - g_{zz} \muB B_{z} \sum_{i} \hat{S}^{z}_{i},
\end{equation}
with $J_1 = 0.326(22)$ meV, where the $g$-tensor component $g_{zz} = 1.89(3)$ is obtained from the gradient of the Larmor mode [Fig.~\ref{fig:hp}(d)]; the LSWT dispersion obtained from this model, marked by the red lines in Figs.~\ref{fig:hp}(e-f), indeed fits extremely well. The model we deduce is consistent with the prediction of $S_{\rm{eff}} = 1/2$ character, minimal anisotropy (other than the $g$-tensor), and predominant antiferromagnetic Heisenberg superexchange in \ce{Yb}$^{3+}$ magnets with edge-sharing octahedral coordination made in Ref.~\cite{RauGingras2018}. 

\begin{figure}[t]
\includegraphics[width=\linewidth]{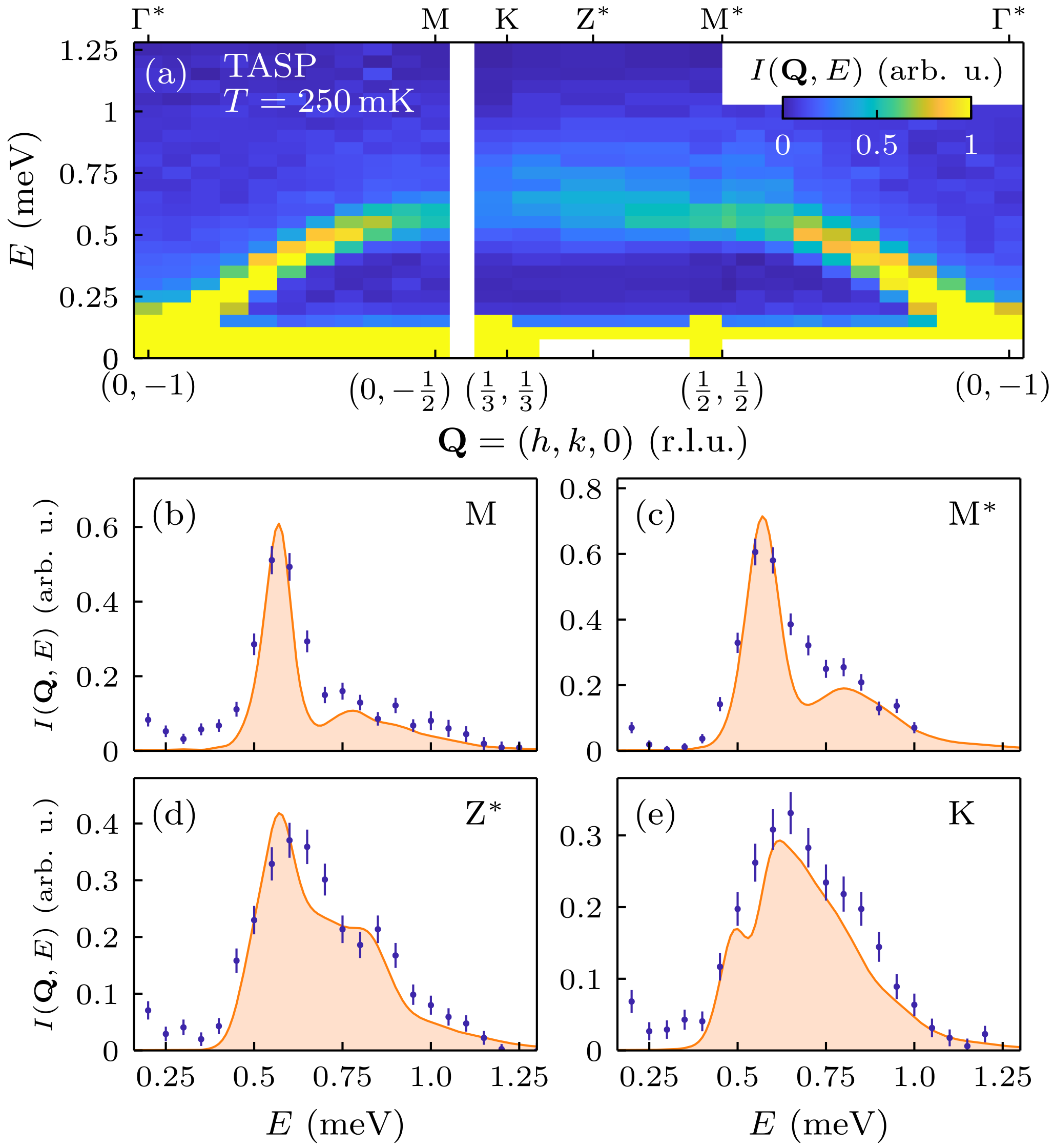}
\caption{\label{fig:zfe} {\bf Zero-field spectrum.} 
(a) INS intensity measured along the paths $\Gamma^*$-M and K-M$^*$-$\Gamma^*$ by INS. (b-e) Quantitative comparison of INS (symbols) and MPS results (beige shading) for constant-${\bf Q}$ spectra measured and computed at M (b), M$^*$ (c), Z$^*$ (d), and K (e).}
\end{figure}

\emph{Intermediate-field spectra.}---We proceed to the regime at all finite fields up to saturation, focusing in Fig.~\ref{fig:Hdep} on the M$^*$ and K points. First at the descriptive level, we draw attention to four field-induced phenomena that are visible in experiment [Figs.~\ref{fig:Hdep}(a,d)] and can be studied both qualitatively [Figs.~\ref{fig:Hdep}(b,e)] and quantitatively [Figs.~\ref{fig:Hdep}(c,f)] by MPS. (i) There is a significant renormalization of the one-magnon branch compared to LSWT, which is strongly downward with increasing field. (ii) The clearest scattering continuum is a broadened ``shadow'' of the one-magnon branch that rises linearly above it with the applied field at $B < \tfrac{2}{3} B_{\rm sat}$, and in fact appears at all $\mathbf{Q}$ \cite{Sala2023}. (iii) The second-strongest continuum is a low-energy feature that develops at higher fields ($B > \tfrac{1}{2} B_{\rm sat}$) and seems to drain weight away from the one-magnon branch. Continua (ii) and (iii) are more complex than the two-magnon continuum obtained from SWT \cite{Sala2023}, which corresponds to two energetically renormalized (at order $1/S$) but noninteracting magnons, and our results at fields $B > {\textstyle \frac12} B_{\rm sat}$ reveal significantly more insight into interaction effects. (iv) At the K point we observe a complete decay of the one-magnon branch as the field is reduced below 1.5 T [Figs.~\ref{fig:Hdep}(d-f)], and below we discuss the apparent zero-field continuum that results; we remark that the energy of the one-magnon branch at M$^{*}$ also shows a strong upward trend from 1.5 to 0 T [Figs.~\ref{fig:Hdep}(a,b)]. We stress again that the quantitative agreement between INS and MPS in Fig.~\ref{fig:Hdep} is achieved with a single scale factor for the intensity (Sec.~S4 of the SM \cite{sm}), confirming qualitatively (1) that the n.n.~HHAF is the correct model for YbBr$_3$ and (2) that all of the observed features are intrinsic to the HHAF.

\begin{figure*}[t]
\includegraphics[width=\linewidth]{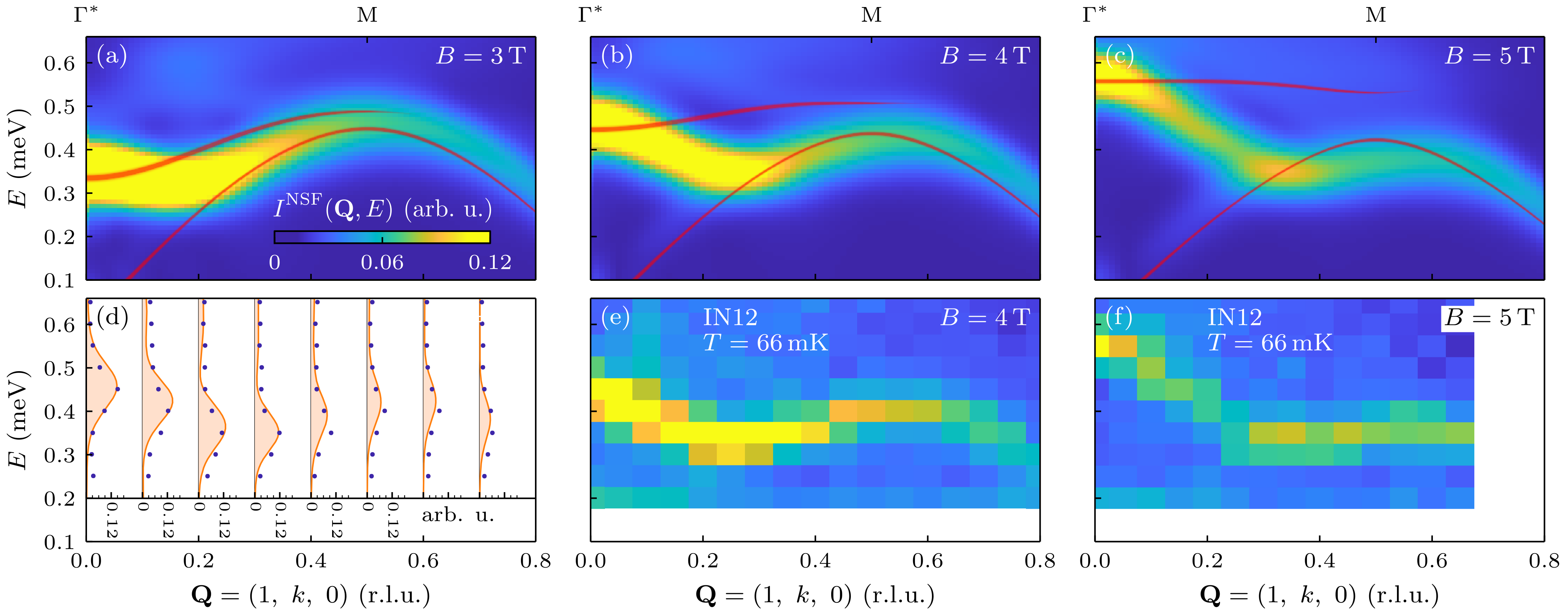}
\caption{\label{fig:Qdep} {\bf Roton-like excitation.} (a-c) MPS calculations of the longitudinal (non-spin-flip) spectral response, $I^{\rm NSF} (\mathbf{Q},E)$, along the $\Gamma^*$-M direction for fields of 3, 4, and 5 T, showing the field-induced evolution of a roton-like feature. Red lines show the magnon branches within LSWT. (d) Constant-$|{\bf Q}|$ cuts through the INS and MPS spectral functions at 4 T for eight wave vectors. Error bars on the INS data are smaller than the symbol sizes. (e-f)
Intensity in the NSF channel measured by polarized INS at $T = 66$ mK, showing the roton-like mode at 4 and 5 T.}
\end{figure*}

Although the field-induced downward renormalization of the one-magnon branch at M$^*$ and K (i) is captured qualitatively at the $1/S$ level \cite{Sala2023}, our demonstration of continuum scattering at all higher energies is consistent with the picture \cite{Verresen2019} of a strong mutual repulsion between this branch and the developing continua, in turn suggesting strong magnon interactions and a multimagnon origin for the latter. However, the dominant effect of these continua, particularly continuum (iii), is the near-complete disappearance of the single magnon branch at fields approaching $B_{\rm sat}$. While this magnon decay could be regarded as a diverging width or as a field-induced fractionalization, its progressive nature and the energetic separation of the magnon from continuum (iii) reinforce the picture of strong two-magnon interactions \cite{Verresen2019}, rather than the abrupt decay at a well defined two-magnon continuum boundary \cite{zc2013,mc2016} characteristic of noninteracting magnons. 

Turning to the low-field continuum (ii), it is easy to see that its energy offset from the one-magnon branch is the Larmor energy, $\hbar \omega_{\rm L}$, which explains its linear rise with the field (Fig.~\ref{fig:Hdep}). This suggests an interpretation as a collective two-magnon state formed from a single magnon at $\mathbf{Q}$ and a Larmor mode at $\Gamma$. We remark that this collective state is remarkably sharp at all wave vectors in the SLHAF, where it merits the name ``shadow mode'' \cite{slsm2024}. In the HHAF we use the term ``magnon shadow'' because both our MPS and INS measurements on \ce{YbCl3} \cite{Sala2023} show that it is considerably broader at most $\mathbf{Q}$ points, illustrating significant differences between these two paradigm unfrustrated quantum spin models. The continuum of two noninteracting magnons \cite{Sala2023} captures the position of the shadow qualitatively, but finds intensities that are more diffuse in energy and are dominated by van Hove points that both INS and our MPS results show to be artificial over a wider field range. This comparison highlights the key signatures of strong magnon-magnon interactions, which are captured by MPS but are missing from SWT at the $1/S$ level. We note here that the field-controlled spectrum changes very rapidly as the field is reduced below $B_{\rm sat}$, reflecting a very low energy scale associated with the linear part of an otherwise quadratic magnon spectrum, and hence a very low threshold for the formation of bound and scattering states [continuum (iii)] contributing to magnon decay in this regime. 

\emph{Zero-field spectrum.}---In Fig.~\ref{fig:zfe}(a) we replot the INS data of Ref.~\cite{Wessler2020} to analyze the previously reported tendency that magnon-like excitations in the inner BZ evolve into excitation continua near the BZ boundary. Our MPS calculations reflect the maximal quantum entanglement at zero field through some persisting weak dependence on the cylinder size and bond dimension, although these differences vanish when the spectrum is convolved with the instrumental resolution (Sec.~S4 of the SM \cite{sm}). Using our MPS results to quantify continuum formation on the zone boundary, at M we find a rather well defined magnon accompanied by only weak continuum scattering at higher energies [Fig.~\ref{fig:zfe}(b)]. At M$^*$, the magnonic peak remains and the continuum is slightly stronger [Fig.~\ref{fig:zfe}(c)]. On proceeding towards K, there is a progressive weakening of the magnon and strengthening of the continuum [Fig.~\ref{fig:zfe}(d)], but only at the K point do we find a near-complete loss of sharp magnon intensity [Fig.~\ref{fig:zfe}(e)]. Our MPS results make clear that this continuum-like K-point spectral function, proposed \cite{Sala2023} as the honeycomb analog of the $(\pi,0)$-anomaly in the SLHAF \cite{Headings2010,Plumb2014,DallaPiazza2014}, is an intrinsic quantum many-body property of the HHAF, as are the continua lying above the putative one-magnon branch for all of the zone-boundary points [Figs.~\ref{fig:zfe}(b-d)]. We remark that the value of $J_1$ extracted from our high-field data continues to provide an optimal fit at zero field.

\emph{Roton-like mode.}---We close by considering the $\mathbf{Q}$-dependence of the spectrum in more detail. The intensity measured by unpolarized INS is a sum of transverse and longitudinal spin excitation channels, which we define relative to the field direction ($\boldsymbol{z}$) and can separate readily in our MPS calculations. In this way we noticed highly unconventional behavior of the longitudinal spectral weight at intermediate fields that is reminiscent of the roton in liquid $^4$He \cite{Landau1941}, with a distinctive dip of the mode energy at small but finite $|\mathbf{Q}|$. As Figs.~\ref{fig:Qdep}(a-c) show, this mode involves an abrupt transfer of spectral weight between the two one-magnon branches of LSWT to produce a sharp magnon-like mode in a part of reciprocal space absolutely not predicted by SWT, even with $1/S$ corrections. 

Motivated by this observation, we performed a polarized INS experiment on the TAS spectrometer IN12 at the ILL, as detailed in Sec.~S2 of the SM \cite{sm}. Equating the longitudinal response directly with the non-spin-flip (NSF) channel, the measured spectra [Figs.~\ref{fig:Qdep}(e-f)] find the ``honeycomb roton'' to be a resolution-limited mode that agrees quantitatively in both position and intensity with the MPS predictions over a range of fields [Fig.~\ref{fig:Qdep}(d)]. Here we have already identified the mutual repulsion of the magnon branch and higher-lying continua, and in this context the field-dependent rotonic minimum in Figs.~\ref{fig:Qdep}(a-c) corresponds well to the changing {\bf Q} position of the minimum in the nearby continuum. Thus the honeycomb and liquid-He rotons share the property that both constitute collective behavior emerging from strong multiparticle interactions that cause strong mutual renormalization of both the lower branch and the higher continuum. We note that this roton-like behavior is also visible in the 4 T INS data on YbCl$_3$ \cite{Sala2023}, but not in the spectrum obtained by nonlinear SWT, pointing again to the effects of strong magnon-magnon interactions that are captured by our MPS calculations. 

\emph{Discussion.}---We reiterate that our INS measurements and MPS calculations show full quantitative equivalence over the entire range of {\bf Q}, $E$, and $B$, confirming both the model and the spectral function. Thus our results reveal multiple phenomena in the HHAF spectrum which are hallmarks of intrinsic many-body interactions that are manifestly strong, although they are not of the type required to produce a QSL. Considering first the elementary excitations, over most of the BZ and field range we find well defined magnonic modes, albeit with significant energetic renormalization that appears to be a consequence of mutual repulsion from extensive higher-lying continua (Fig.~\ref{fig:Hdep}) \cite{Verresen2019}. The exception is around the K point at low fields, where it is hard to argue for a true quasiparticle [Figs.~\ref{fig:Hdep}(e) and \ref{fig:zfe}(f)]. As in the SLHAF at $(\pi,0)$, one anticipates an interpretation of this feature in terms of multimagnon states \cite{Powalski2016,Powalski2018} or of magnon fractionalization \cite{Shao2017,Yu2018,Zhang2022}. Two-magnon binding has been found in the square lattice in a field \cite{slsm2024} and a further example of multimagnon resonances would be the correlated plaquette fluctuations proposed in Ref.~\cite{Wessler2020}. Fractional quasiparticles are expected around deconfined quantum critical points (DQCPs), and extensive studies of the frustrated HHAF interactions \cite{Fouet2001,Mulder2010, Oitmaa2011,Farnell2011, Albuquerque2011,Ganesh2013,Zhu2013,Gong2013,Ghorbani2016} show that $J_2$ causes the spectrum to soften at K, where N\'eel order gives way to a plaquette (hexagon) valence-bond crystal at a DQCP. Possible fingerprints of this deconfinement have been reported \cite{Ferrari2020,Gu2022} at finite energies even for $J_2 = 0$, and our demonstration that a magnonic excitation forms above 1.5~T [Fig.~\ref{fig:Hdep}(e)] would be consistent with a field-induced spinon reconfinement.

One of our signature findings is field-induced magnon decay, which we show to be a progressive phenomenon following the strong-interaction scenario \cite{Verresen2019} where spectral weight shifts systematically to higher-lying continua, with no sharp onset in field or wave vector. In fact the magnons never disappear completely, even if the vast majority of their weight has moved to higher energies near saturation (Fig.~\ref{fig:Hdep}). In this context, the coherent collective modes we observe come with a caveat: the magnon shadow forming at low fields (Fig.~\ref{fig:Hdep}, \cite{Sala2023}) still has continuum rather than single-mode character, not least because the magnons themselves become broadened entities near the BZ boundary. Where the magnon branches are largely conventional, in the inner BZ and at lower fields, then we find a highly unconventional shift of spectral weight between them, and this is the roton-like excitation (Fig.~\ref{fig:Qdep}).  

\emph{Conclusions.}---We have measured and computed the spectral function of the honeycomb antiferromagnet \ce{YbBr3} at all fields from zero to saturation. We use the spectra in the fully polarized phase to establish that the spin Hamiltonian is of pure Heisenberg and nearest-neighbor form. We find a rich and complex field-induced spectral response that deviates from spin-wave predictions in many regards. At low energies, magnonic modes appear at unexpected positions, influenced by repulsive interactions with nearby excitation continua. At low fields, quantum fluctuations destroy the magnon around the K point. At higher fields, the magnons are broadened by their interactions to the threshold of complete decay. However, interaction effects are not only destructive, but also create coherent features over significant regions of the spectrum. These include collective magnon shadows, where the magnon branches interact strongly with the Larmor mode, and roton-like modes, where the scattered intensity switches abruptly between magnonic branches to produce a strong and sharp excitation with non-monotonic dispersion. We conclude that a wealth of interacting many-body phenomena remain to be understood in the complete field-induced spectral functions of quantum magnetic systems, even in the absence of frustration.

\emph{Acknowledgments.}---We are grateful to {\O}.~S.~Fjellv{\aa}g for technical assistance and S.~Nikitin for helpful discussions. We acknowledge the financial support of the Swiss National Science Foundation (SNF) under Grant No.~200020\_172659. We thank the Paul Scherrer Institute for the allocation of neutron beam time on the instruments CAMEA, TASP, and ZEBRA. Beam time on IN12 at the Institut Laue Langevin was supported by the Swiss State Secretariat for Education, Research and Innovation through a CRG grant.

\emph{Data availability.}---INS data collected at the ILL in the course of the present work are available as Ref.~\cite{ill}.

%\bibliography{ybbr3_biblio}
%

\onecolumngrid

\phantom{x}

\clearpage

\setcounter{figure}{0}
\renewcommand{\thefigure}{S\arabic{figure}}

\setcounter{section}{0}
\renewcommand{\thesection}{S\arabic{section}}

\setcounter{equation}{0}
\renewcommand{\theequation}{S\arabic{equation}}

\setcounter{table}{0}
\renewcommand{\thetable}{S\arabic{table}}

\onecolumngrid

\vskip 12mm

\noindent
{\large{\bf {Supplemental Materials to accompany the manuscript}}}

\vskip 2mm

\noindent
{\large{\bf {Field-Induced Magnon Decay, Magnon Shadows, and Roton-like Excitations in the Honeycomb}}}

\noindent
{\large{\bf {Antiferromagnet YbBr$_3$}}}

\vskip 2mm

\noindent
J. A. Hernández, A. A. Eberharter, M. Schuler, J. Lass, D. G. Mazzone, R. Sibille, S. Raymond, K. W. Kr\"amer, B. Normand, B. Roessli, A. M. L\"auchli, and M. Kenzelmann

\vskip 8mm

\twocolumngrid

\section{Crystal growth, characterization and preparation}\label{sec:supp_crystal}

The samples of \ce{YbBr3} used in this work were the same crystals studied in Ref.~\cite{Wessler2020}, which were grown by the Bridgman method. YbBr$_3$ was synthesized from Yb$_2$O$_3$ (99.999\%) by the NH$_4$Br method and purified by sublimation as described in Ref.~\cite{Kraemer1999}. From unpolarized neutron diffraction data on powder and single-crystal samples of \ce{YbBr3}, we found the lattice parameters at room temperature to be $a = b = 6.976$ $\rm{\AA}$ and $c = 19.115$ $\rm{\AA}$, and the structure to be consistent with the BiI$_3$-type structure, and space group $R\Bar{3}\;(148)$, as determined in Refs.~\cite{BiI3a,BiI3b}. We remark that the Yb$^{3+}$ ions in YbBr$_3$ are located on an undistorted honeycomb lattice; by contrast, the closely related material YbCl$_3$ crystallizes in the AlCl$_3$-type structure with space group $C2/m\;(12)$ \cite{AlCl3}, whose monoclinic symmetry means that the Yb$^{3+}$ ions in YbCl$_3$ lie on a slightly distorted honeycomb lattice.

Because \ce{YbBr3} is a highly hygroscopic material, all sample handling was performed in dry and \ce{O}-free conditions. Two single crystals were selected for neutron scattering. For all of our inelastic experiments, we used a large crystal of 5.33 g, which was sealed under He in a gas-tight \ce{Cu} container in order to guarantee proper thermal contact between the crystal and the dilution refrigerator used at low temperatures. For our neutron diffraction experiment, a small crystal of 0.54 g was prepared inside the same type of \ce{Cu} container.

\begin{figure*}[t]
\includegraphics[width=\linewidth]{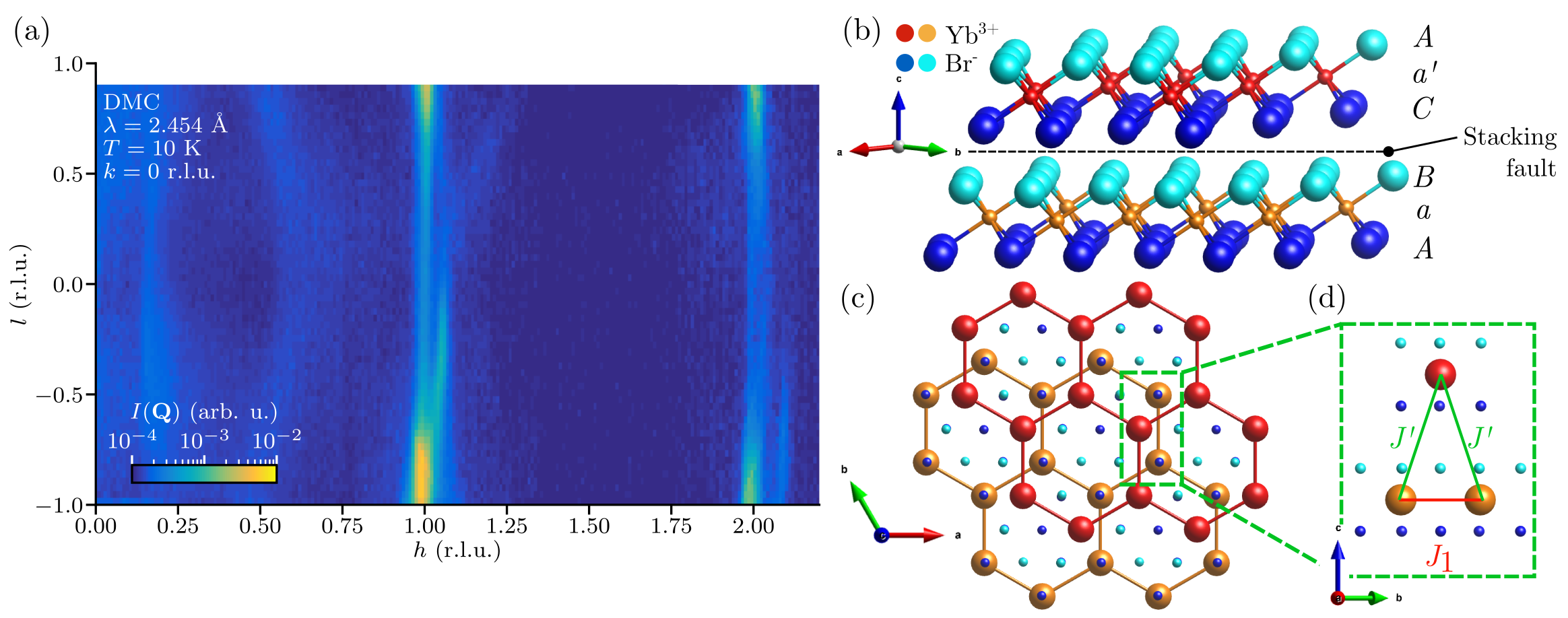}
\caption{{\bf Nuclear scattering and stacking faults in \ce{YbBr3}.} (a) Unpolarized neutron diffraction data measured in the paramagnetic phase and shown for the $(h0l)$ scattering plane. The modulated rods of nuclear intensity reveal the lack of structural correlations along the stacking direction. (b) Representation of the faulted stacking sequence AaB-Ca$^\prime$A. (c) Projection of the two layers around this fault on the $(\boldsymbol{ab})$ plane. The Yb$^{3+}$ ions in the honeycomb a (a$^\prime$) layer are shown in orange (red) and the A(B)-layer Br$^-$ ions in dark (light) blue. The stacking-fault translation vector taking AaB to Ca$^\prime$A is $\tfrac{2}{3}\boldsymbol{a}+\tfrac{2}{3}\boldsymbol{b}+\tfrac{1}{3}\boldsymbol{c}$, which places all the Yb$^{3+}$ ions in layer a$^\prime$ equidistant from two nearest-neighbor Yb$^{3+}$ ions in layer a, as shown in the projection on the plane perpendicular to $(\boldsymbol{ac})$ in panel (d). The Yb-Yb distances for the $J_1$ and $J^\prime$ bonds in panel (d) are $r(J_1) = 4.028$ \AA~and $r(J^\prime) = 6.783$ \AA. In panel (b) we show the Yb$^{3+}$ and Br$^-$ ions with the correct ratio of their ionic radii, whereas in panels (c) and (d) we have reduced the Br$^-$ radii for a clearer visualization of the honeycomb planes and interlayer geometry. }
\label{sup:rods}
\end{figure*}

The symmetry $R\bar{3}$ admits only nuclear Bragg reflections whose indices satisfy $- h + k + l = 3n$, with $n$ an integer or zero. However, \ce{YbBr3} is prone to develop crystalline stacking faults between honeycomb planes due to the weak van der Waals forces that couple these. This faulting lowers the space-group symmetry and would result in the observation of diffracted intensity at forbidden wavevectors. To gauge the extent of faulting in \ce{YbBr3}, we performed an unpolarized neutron diffraction measurement on a single-crystal sample using the DMC diffractometer at PSI, working at a temperature ($T = 10$ K) well above that of any 3D or 2D magnetic correlations. As Fig.~\ref{sup:rods}(a) shows clearly, we find nuclear intensity at the forbidden positions $\mathbf{Q} = (1, 0, 0)$ and $(2, 0, 0)$ r.l.u., which takes the form of rods that span the entire Brillouin zone along the $\boldsymbol{c}^{*}$ axis. These results reflect very extensive interlayer faulting in the nuclear structure of \ce{YbBr3}, which can be expected to have clear consequences for any interlayer magnetic correlations developing at low temperatures. 

To interpret the Bragg rods, we first consider the nature of a stacking fault. The halide ions in unfaulted \ce{YbBr3} are hexagonally close-packed (HCP), meaning that they have an ABABAB sequence of triangular layers; between AB pairs, the Yb$^{3+}$ ions occupy honeycomb layers (i.e.~they fill 2/3 of the available sites) with effective abcabc stacking, where the small letters signify that the sequence refers only to the metallic ions (neglecting the halide ions). Between BA pairs of the halide stack, all metal sites are empty, and this is where a stacking fault occurs, as represented in Fig.~\ref{sup:rods}(b): the halide sequence is then $\dots$ABAB-CACA$\dots$CACA-BCBC$\dots$, where - denotes a fault and there is only one possibility, because AB-BC (or cyclic permutations) is not permitted. Thus a stacking fault is not at all random in the honeycomb plane, because there is only one type of fault and it does not alter the axes of the honeycomb planes, ensuring our ability to measure a coherent 2D magnetic excitation spectrum in the presence of extensive faulting. 

Quantitatively, the intensities within the $L$-axis Bragg rods in Fig.~\ref{sup:rods}(a) differ by a factor of 40, a result that can be reproduced by assuming a 5\% concentration of stacking faults, which implies an average fault spacing of order 10 nm. Thus while the only random aspect of the structure is the stacking sequence of halide bilayers, the crystalline axes being fixed and the positions of the halide ions having only the three possibilities A, B, and C, the presence of 10$^6$ stacking faults in a cm-sized crystal sets a clear limit on the positional correlations along the $\boldsymbol{c}$ axis. 

Turning to the locations of the metallic ions, their interlayer translation vector is changed by a stacking fault from $\tfrac{1}{3}\boldsymbol{a}+\tfrac{2}{3}\boldsymbol{b}+\tfrac{1}{3}\boldsymbol{c}$ to $\tfrac{2}{3}\boldsymbol{a}+\tfrac{2}{3}\boldsymbol{b}+\tfrac{1}{3}\boldsymbol{c}$, as we represent in Fig.~\ref{sup:rods}(b). This maps the a layer to a set of locations not in the abc sequence at all, which we denote in Fig.~\ref{sup:rods} by a$^\prime$. The ideal nature of the HCP halide stack ensures that every Yb$^{3+}$ ion in layer a$^\prime$ is equidistant from two such ions in layer a on the opposite side of the fault, as shown in Figs.~\ref{sup:rods}(c,d). This situation ensures complete magnetic frustration of all the bonds across a stacking fault, which guarantees a total loss of magnetic correlations in the $\boldsymbol{c}$ direction in YbBr$_3$. 

Here we draw attention to the contrasting situation in YbCl$_3$, where the different structure and monoclinic space group create an aaaa stacking of the Yb$^{3+}$ ions. In both materials, one would expect by the Goodenough-Kanamori rules that the interlayer Yb-X-X-Yb path, which possesses no special geometric properties, has a very weak and antiferromagnetic superexchange interaction. Thus the effectively ferromagnetic long-ranged order in YbCl$_3$ \cite{Sala2021} implies that the aa interplane bond is not the strongest one in the system, and acts to create the partial frustration of a different interplane interaction. However, the monoclinic structure ensures that this frustration is not complete in the same way as at a stacking fault in YbBr$_3$, where perfect frustration is enforced by the geometry.  

\section{Magnetic neutron scattering}\label{sec:supp_neutronexp}

\subsection{Unpolarized neutron diffraction}

For an independent estimate of the saturation field of \ce{YbBr3}, we performed a diffraction experiment on the thermal neutron diffractometer ZEBRA at PSI with the field applied along the crystallographic $c$ axis. The combination of a dilution refrigerator and a superconducting $15$-Tesla cryomagnet allowed us to maintain the sample at a temperature $T = 55$ $\rm{mK}$ and under an applied magnetic field of up to $B_{z} = 11$ T. The experiment was performed with an incident neutron wavelength $\lambda = 1.178$ $\rm{\AA}$ and without additional beam collimators in order to maximize the scattered neutron intensity. At $\mathbf{Q} = \Gamma = (1,1,0)$, this intensity is the sum of a field-independent structural Bragg peak and a field-induced magnetic component, which in turn is the sum of static and dynamic magnetic correlations. The former are proportional to the square of the field-induced ordered moment along the field direction while the latter are given by the field-induced magnetic excitations. Because we did not use an analyzer crystal, and because of the short neutron wavelength, all of the dynamic correlations were integrated and appear as contributions to the measured neutron intensity \cite{Collins1989}. At the saturation field, $B_{\rm sat}$, the magnetic contribution to the scattering intensity also saturates, at a maximum value corresponding to the fully polarized spin configuration. The purely magnetic scattering intensity is shown as a function of the applied field in Fig.~1(c) of the main text.

\subsection{Unpolarized INS}

\subsubsection{Field-polarized phase}

The magnon excitations of \ce{YbBr3} were measured in the fully field-polarized phase using the cold-neutron multiplexing spectrometer CAMEA at PSI. A superconducting $11$-Tesla cryomagnet and a dilution insert allowed the sample temperature to be reduced to $T = 50$ mK and the magnetic field to be set to $B_{z} = 9.73$ T. We operated the spectrometer with an incident neutron energy of $E_i = 5$ meV, thereby probing large areas of reciprocal space simultaneously up to a maximum energy transfer of 1.8 meV and with a resolution at the elastic position of 0.21 meV. A combined filter-collimator system consisting of a cooled Be filter and a radial Soller collimator was placed between the sample position and the analyzer array, the former suppressing higher-order Bragg contamination from the monochromator and the latter eliminating scattering from the sample environment.

\begin{figure}[t]
\includegraphics[width=\linewidth]{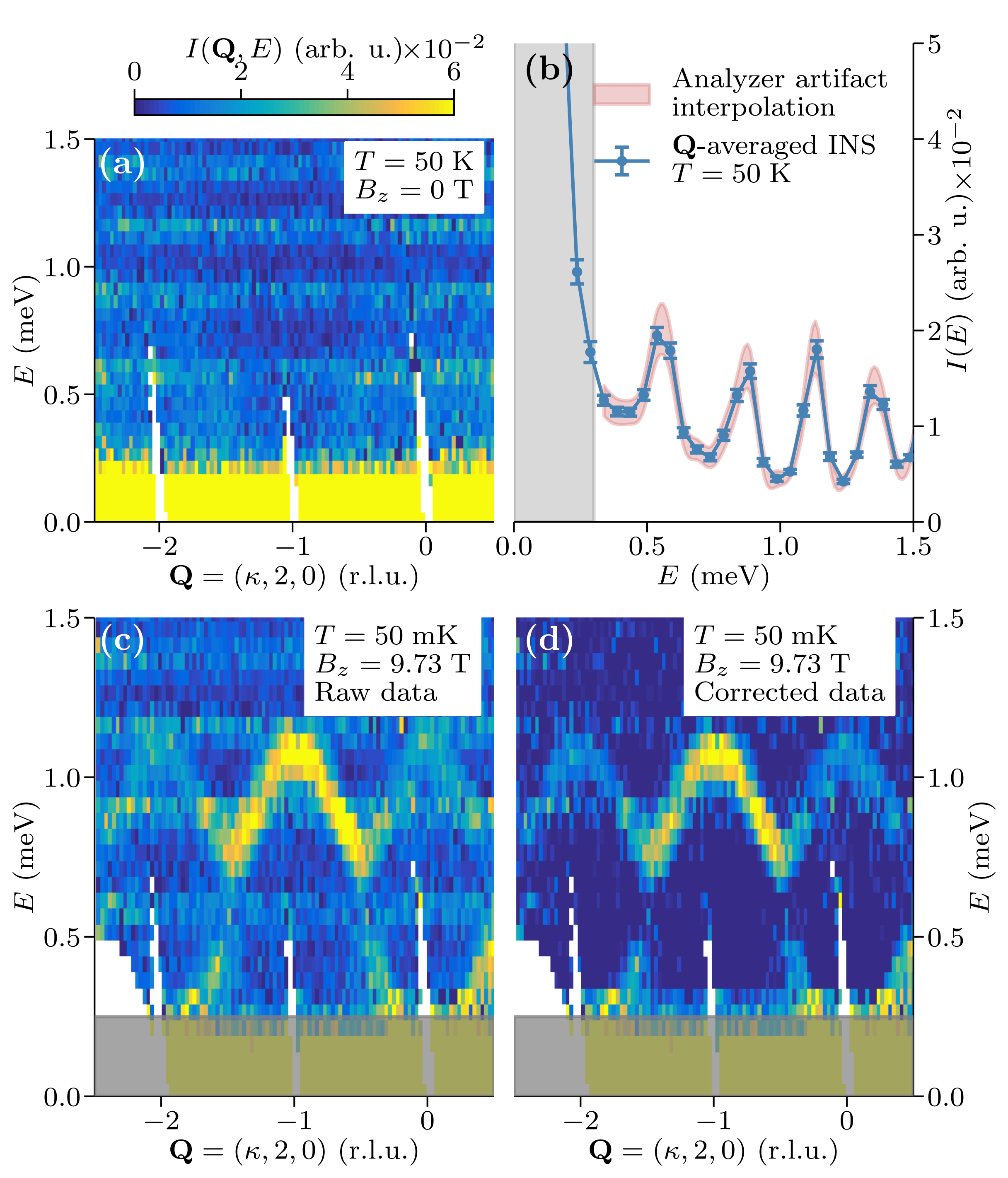}
\caption{{\bf Data correction method on CAMEA.}
(a) INS data in the paramagnetic phase, with $\mathbf{Q}$-independent intensity oscillations arising due to the finite number of analyzer crystals.
(b) $\mathbf{Q}$-averaged intensity and propagated $1\sigma$ error bar on the intensity oscillation. 
(c) Raw and (d) corrected INS data collected along the $\mathbf{Q} = (\kappa, 2, 0)$ direction in the fully-polarized phase at base temperature. Panel (d) shows clearly the two resolution-limited magnon branches. Gray-shaded regions indicate energy ranges dominated by incoherent scattering, which were neglected for data-correction purposes.}
\label{sup:bgsub}
\end{figure}

The reduction of all CAMEA data was performed using the \textsc{MJOLNIR} software \cite{Lass2020}. Datasets were corrected for spurious Bragg tails that extend to finite energies by direct masking of the affected $(\mathbf{Q}, E)$ points. To better discern the magnon dispersion at base temperature, one zero-field dataset was measured at $T = 50$ K, well in the paramagnetic phase of the system. This dataset was used to subtract instrumental artifacts arising from the analyzer array, which take the form of the oscillations in intensity as a function of energy transfer visible in Fig.~\ref{sup:bgsub}(a). Because the intensity at 50 K is independent of $\mathbf{Q}$, integration over $\mathbf{Q}$ provides an accurate measure of the intensity contribution due to the analyzer artifact [Fig.~\ref{sup:bgsub}(b)]. Interpolation of these data and propagation of the $1\sigma$ uncertainty were used to create an effective model for this instrumental artifact, which is independent of field and temperature, and hence to subtract it directly from the data at $T = 50$ mK at $B_{z} = 9.73$ T. Figures \ref{sup:bgsub}(c-d) illustrate the results of this procedure, and similar procedures are reported in Refs.~\cite{Facheris2022,Sala2023,Lass2024}. In this way we extracted the field-polarized magnon dispersion at 186 distinct $\mathbf{Q}$ points along the two inequivalent high-symmetry directions of the reciprocal lattice. In Sec.~\ref{sec:supp_lswt} we describe how this dispersion was used to determine the effective low-energy spin Hamiltonian of \ce{YbBr3}.

\subsubsection{Field-canted phase}

We investigated the field-dependence of the magnetic excitations at all fields up to saturation at the cold-neutron spectrometer TASP at PSI. Operating in fixed-$k_f$ mode with $k_{f} = 1.3$ $\rm{\AA}^{-1}$ resulted in an energy resolution of 0.079 meV FWHM at the elastic position and a dilution refrigerator was used to set a base temperature of $T = 100$ mK. This experiment was used to measure the spectra at fixed momentum transfer for the different applied fields shown in Fig.~2 of the main text. 

\begin{figure}[t]
\includegraphics[width=\linewidth]{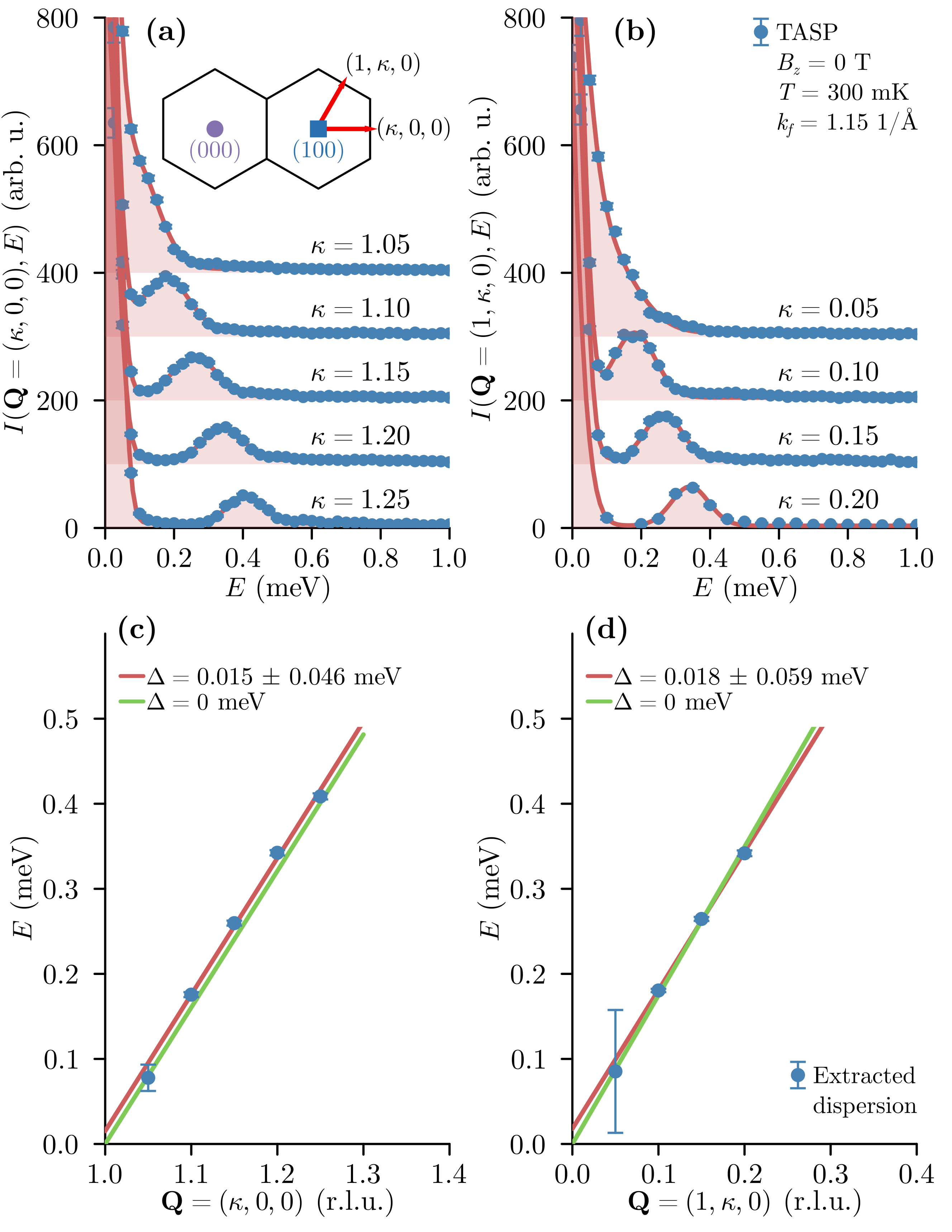}
\caption{ {\bf Zero-field magnetic excitations of \ce{YbBr3} at low momentum transfer.}
INS scans in high-resolution mode performed around the $\Gamma^{*}$ point along the high-symmetry directions $(\kappa, 0, 0)$ (a) and $(1, \kappa, 0)$ (b).
Spectra at different momenta are offset vertically by $+100$ units for clarity. Error bars on the data points, marking $1\sigma$ standard deviations, are smaller than the symbol size. The inset in panel (a) denotes the two high-symmetry directions in the reciprocal lattice. The solid line is a fit to the sum of two Gaussian profiles that accounts for both the inelastic signal and the incoherent scattering at the elastic position. The dispersion of a magnon-like excitation is extracted from the fitted peak position of the inelastic signal in $E$.
(c-d) Dispersion of magnon-like excitations at low momentum transfer. The red line is a fit to the form $E = c|\kappa| + \Delta$, from which we estimate a spin gap, $\Delta$. Because the dispersions extrapolate to zero at $|\mathbf{Q}| = 0$ within statistical and experimental error, the dispersion data are described equally well described by a one-parameter fit with $\Delta = 0$ (green lines).}
\label{sup:gap}
\end{figure}

\begin{figure*}[t]
\includegraphics[width=\linewidth, keepaspectratio]{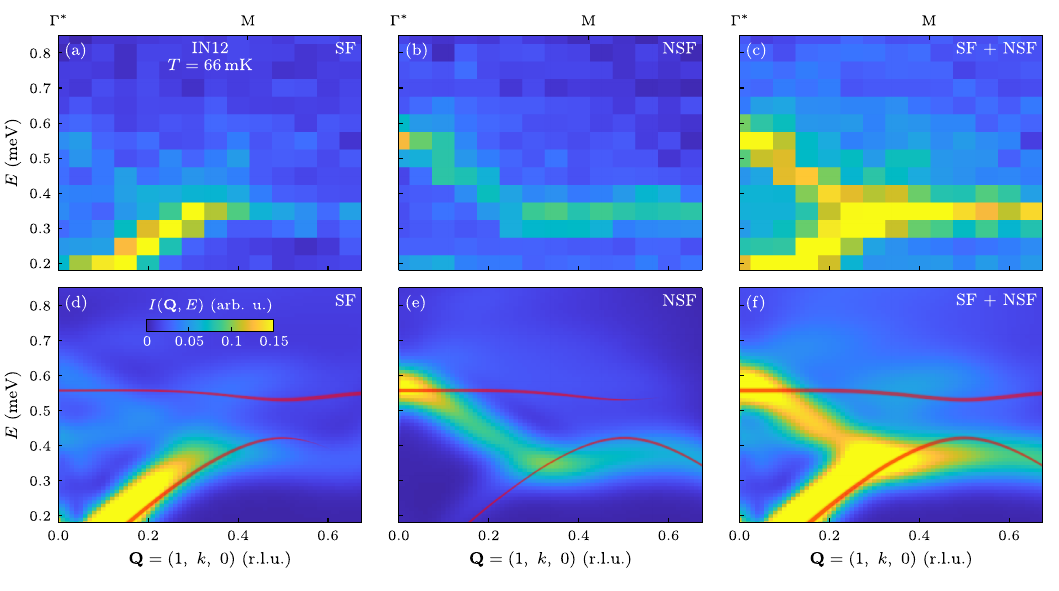}
\caption{{\bf Longitudinal and transverse spin excitation spectra.} (a-c) Polarized neutron intensities measured on IN12 at 5 T. The intrinsic (corrected and normalized) INS intensities in the transverse, or spin-flip (SF), and longitudinal, or non-spin-flip (NSF), channels are shown in panels (a) and (b), respectively, and their sum in panel (c). The red lines show the calculated magnon dispersion for the two magnon branches of LSWT evaluated using Eq.~(1) of the main text with $J = 0.326$ meV, and their width is proportional to the calculated intensity. (d-f) MPS calculation of the SF (d), NSF (e), and summed (f) intensities performed for the HHAF with $J = 0.326$ meV at a field equivalent to 5 T, as described in Sec.~\ref{sup:MPS}C.}
\label{sup:in12_5T}
\end{figure*}

\subsubsection{Zero field}

We have re-analyzed the zero-field INS spectra reported in Ref.~\cite{Wessler2020} by comparing the data to our MPS calculations, as shown in Fig.~3 of the main text. These spectra were collected at TASP at a temperature of $T = 250$ mK, using a fixed $k_{f}$ of 1.3 $\rm{\AA}^{-1}$ and no additional beam collimation such that the resolution was reported to be 0.08 meV FWHM. In separate measurements, the INS spectrum was investigated at low momentum transfer, meaning close to the $\mathbf{Q} = \Gamma^{*}$ point, with the finest energy resolution possible at TASP to determine whether the magnetic excitations possess a gap in their dispersion or show a lifting of the degeneracy between the two magnon modes. For this experiment, the spectrometer was operated at $T = 300$ mK with a fixed $k_f$ of 1.15 $\rm{\AA}^{-1}$, resulting in an enhanced energy resolution of 0.062 meV FWHM at the elastic position. The monochromator crystal was vertically focusing and the analyzer horizontally focused, again with no beam collimation to maximize the INS intensity. As shown in Figs.~\ref{sup:gap}(a-b), the magnetic excitations at low momentum transfer appeared as well defined peaks. Figures \ref{sup:gap}(c-d) show that their positions in energy extrapolate to zero at $\mathbf{Q} = \Gamma^{*}$ within experimental error, indicating either a vanishing spin gap or one far below 0.062 meV. Thus our MPS spectra are compatible with zero gap. We note that resolving a finite gap in this low energy range would require high-resolution backscattering experiments with long-wavelength neutrons or neutron spin-echo techniques. 

\subsection{Polarized INS}

To study the field-dependence of the longitudinal and transverse spin excitations separately, we performed an INS experiment with polarized neutrons on the IN12 cold-neutron TAS at the ILL \cite{Schmalzl2016}. As noted in the main text, the geometry of our experiment is that the magnetic field is applied along the $\boldsymbol{z}$ axis, which is perpendicular to the honeycomb layers and to the scattering plane ($\boldsymbol{x}\boldsymbol{y}$). This matches with the standard coordinate system used in longitudinal polarization analysis \cite{moon1969, Boothroyd2020}, where the spin polarization of the neutron beam incident upon the sample, $\mathbf{P}_{i}$, is perpendicular to the scattering plane and hence the spin excitations transverse to the applied field are measured in the spin-flip (SF) polarization channel, $\mathbf{P}_{i} \parallel - \boldsymbol{z}$, whereas the longitudinal excitations are contained in the non-spin-flip channel (NSF), $\mathbf{P}_{i} \parallel \boldsymbol{z}$.

In the experiment, the polarization of the incident neutron beam is prepared by a transmission polarizing cavity located upstream from the instrument but after the velocity selector. Static guide fields to maintain the polarization are installed all along the neutron path, including around the double-focusing pyrolitic graphite monochromator. The polarization of the incident beam was controlled with a Mezei spin flipper placed in the neutron guide upstream from the monochromator. The separation of SF and NSF channels is possible due to the Heusler$_{(111)}$ analyzer of the instrument, which serves to select both the energy of the scattered neutron beam and its polarization state. Only energy-analyzed neutrons with final (scattered) polarization $\mathbf{P}_{f} \parallel \boldsymbol{z}$ arrive at the detector. To suppress cross-channel contamination under finite applied magnetic fields, the flipper currents were calibrated in order to maximize the flipping ratio, $R = I_{\text{NSF}}/I_{\text{SF}}$, up to a maximum energy transfer of 1.8 meV at 4 and at 5 T using a standard graphite sample.

The spectrometer was operated in fixed-$k_{f}$ mode with $k_{f} = 1.3$ $\rm{\AA}^{-1}$, resulting in an energy resolution of 0.074 meV at the elastic position, and the sample temperature reached was $T = 66$ mK. We determined the flipping ratio to be $R = 20.06$ by measuring at the $\mathbf{Q} = (1,1,0)$ structural peak of \ce{YbBr3} in zero field. The measured INS intensity was corrected for the efficiency of neutron polarization by assuming an ideal flipper element. Cross-channel contamination between the SF and NSF polarization channels was eliminated by calculating the intrinsic INS intensity in each polarization channel, $I^{\text{SF}}$ and $I^{\text{NSF}}$, from the observed intensities, $O^{\text{SF}}$ and $O^{\text{NSF}}$, using the expression \cite{Wildes2006}
\begin{equation*}
    \begin{bmatrix}
        I^{\text{SF}}\\
        I^{\text{NSF}}\\
    \end{bmatrix} = 
    \frac{1}{2P}
    \begin{bmatrix}
        (P+1) & (P-1)\\
        (P-1) & (P+1)\\
    \end{bmatrix}
    \begin{bmatrix}
        O^{\text{SF}}\\
        O^{\text{NSF}}\\
    \end{bmatrix},
\end{equation*}
where the polarization efficiency, $P$, is determined from the flipping ratio by $P = (R - 1)/(R + 1)$.
For our measured value of the flipping ratio, the polarization efficiency is $P = 90.50\%$, meaning a cross-channel contamination of less than $10\%$. 

\begin{figure}[t]
\includegraphics[width=0.96\linewidth, keepaspectratio]{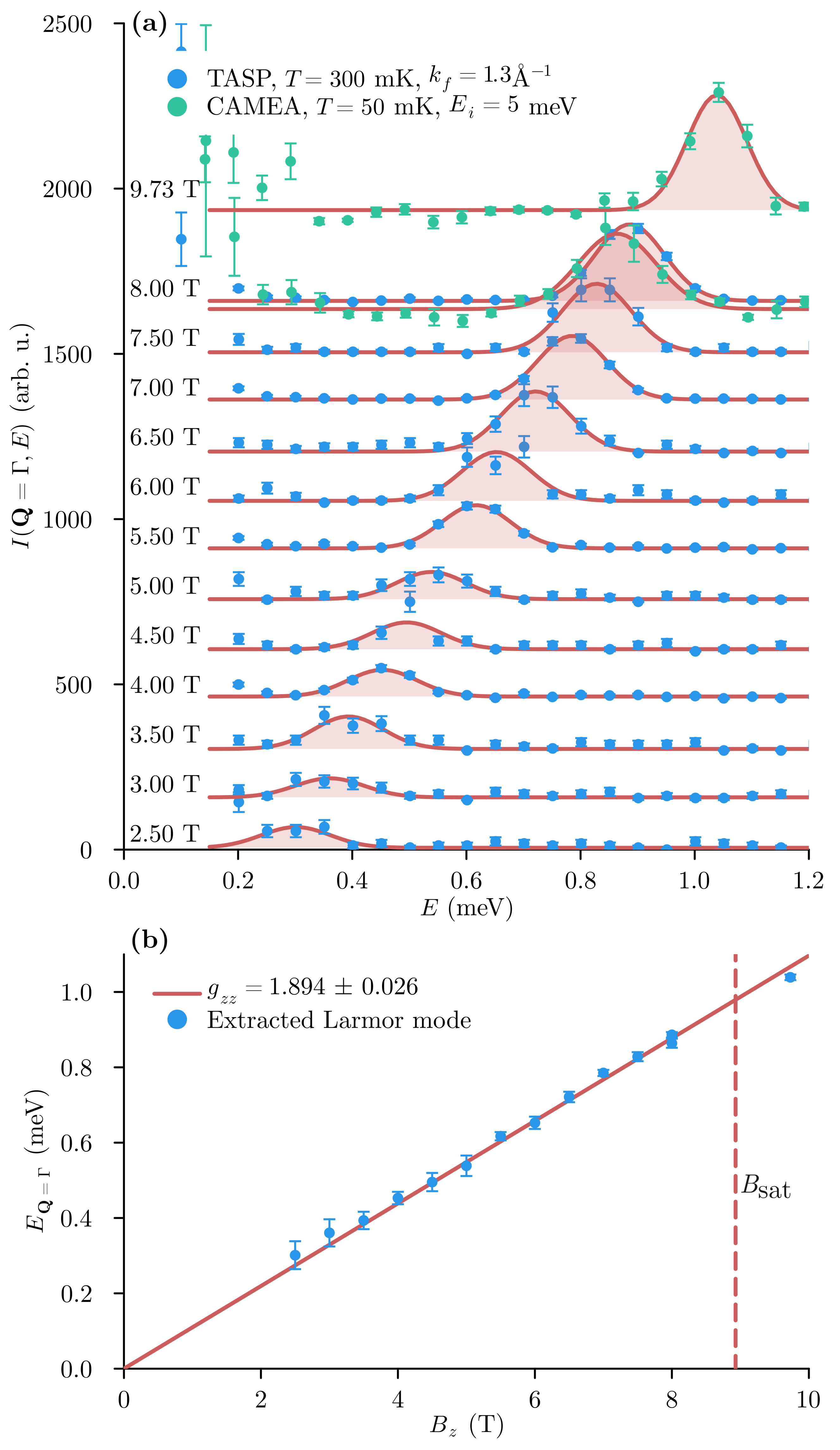}
\caption{ {\bf Field-dependence of the Larmor mode.}
(a) INS spectra measured at the $\Gamma$ point for a sequence of magnetic-field values. All data were collected on TASP, other than two CAMEA datasets taken at $B_{z} = 8$ and 9.73 T; datasets from the different spectrometers were normalized such that the integrated inelastic intensity of the Larmor mode at 8 T is the same. The field-dependence of the Larmor mode was extracted from a fit to Gaussian profile with a fixed width set to the energy resolution of the spectrometers determined at the elastic position. Spectra at different applied fields are offset vertically by $+150$ units for clarity.
(b) Field-dependence of the Larmor mode extracted from the centers of the Gaussians in panel (a). The slope of the Larmor mode (red line) is given by the $zz$-component of the $g$-tensor, which we determine precisely from $g_{zz}\muB$, with $\muB = 0.0579$ meV/T, as $g_{zz} = 1.894(26)$.}
\label{sup:gzz}
\end{figure}

In Fig.~\ref{sup:in12_5T} we demonstrate the separation of the two spin polarization channels by showing the intrinsic INS intensities of the SF [Fig.~\ref{sup:in12_5T}(a)] and NSF [Fig.~\ref{sup:in12_5T}(b), also Fig.~4(e) of the main text] channels measured at $B_{z} = 5$ T. Only the NSF channel displays the remarkable transfer of spectral weight from the optic to the acoustic branch of LSWT part-way between the $\Gamma^{*}$ and M points, which constitutes the honeycomb roton. This is then reflected in the sum of the two channels, which is the unpolarized response [Fig.~\ref{sup:in12_5T}(c)]. In our IN12 experiment, the SF and NSF channels were measured at $B_{z} = 5$ T and the NSF channel was measured at $B_{z} = 4$ T. The NSF intensities measured at 4 and 5 T, and the corresponding MPS calculations, are shown in Fig.~4 of the main text.

\section{Linear spin-wave calculations}
\label{sec:supp_lswt}

\subsection{LSWT for the $S = 1/2$ HHAF}

Linear spin-wave theory (LSWT) is a widely used benchmark method for calculating the excitations of ordered magnetic systems. Formulated in the semiclassical limit of large spin, its accuracy in describing the spectrum of a quantum magnet decreases with increasing quantum fluctuations, although these can be captured in part within a $1/S$ expansion. As a result, while it is no surprise that LSWT fails to describe the physics of ``highly quantum'' ($S = 1/2$, 2D, low-connectivity) spin models, it is customary to use it as an initial guide to the eye that helps to benchmark the anomalous character of a spectral function. It is in this spirit that we used conventional LSWT, in the form of the \textsc{Sunny.jl} package \cite{sunny}, to draw the red dispersion and intensity lines in Figs.~1(d-f), 2(b,e), and 4(a-c) of the main text. We remark that our LSWT dispersions and intensities match quantitatively with the results obtained by a random-phase approximation in Ref.~\cite{Wessler2020}.

\subsection{Determination of the spin Hamiltonian}\label{supp:hamiltonianfit}

In the fully field-polarized phase of a Heisenberg magnet, quantum fluctuations are completely suppressed and the only contribution to the spectrum consists of transverse spin-wave excitations that are described exactly by LSWT. This provides a unique opportunity in quantum magnetic materials with experimentally accessible saturation fields to extract the microscopic parameters of the spin Hamiltonian by combining INS data in the field-polarized phase with LSWT calculations. Here we describe the analysis leading to our determination of the Hamiltonian shown in Eq.~(1) of the main text as the best effective description for the low-energy physics in \ce{YbBr3}.

We start by excluding the possibility of an axial single-ion anisotropy. Such a term is irrelevant in a $S_{\rm{eff}} = 1/2$ system, and the question of whether the low-energy physics of \ce{YbBr3} is appropriately described in the manifold of a single Kramers doublet at low temperatures was addressed in Ref.~\cite{Wessler2022}, where the energy difference between the ground-state manifold and the first excited crystal electric field (CEF) level of the \ce{Yb}$^{3+}$ ions was shown to be approximately 15 meV. In INS, any departure from pure pseudospin-1/2 behavior should be manifest as a gap in the spin-wave spectrum and a lifting of the zero-field degeneracy between the two magnon branches. As Fig.~\ref{sup:gap} makes clear, neither a gap nor a second branch can be discerned in INS measurements performed at the best energy resolution available on TASP, and hence we proceed with a model for the spin dynamics of the system that admits no single-ion anisotropy \cite{Wessler2020,Wessler2022}.

\begin{table*}[t]
    \caption{
        Best-fit Hamiltonian parameters, quoted in meV other than the XXZ anisotropy parameter, $\Delta$, which is dimensionless.
        The quoted errors correspond to the $1\sigma$ uncertainty around the optimum solution.
        $\chi^{2}$ is the goodness of fit.}
    \setlength{\linewidth}{\linewidth}% Shrink \tabcolsep by 30%
    \centering
    \begin{tabular}{ *{6}{c} }
    \toprule
    Model & $J_{1}$ & $J_{2}$ & $J_{3}$ & $\Delta$ & $\chi^{2}$ \\
    \midrule
    $\; J_1$-$J_2$-$J_3$-$\Delta \;$  &   $\; 0.285(21) \;$       &   $\; 0.014(7) \;$   &   $\; -0.008(16) \;$  &   $\; 0.036(51) \;$   &   $\; 0.5255 \;$ \\
    $J_1$-$J_2$-$\Delta$        &   $0.297(31)$       &   $0.007(8)$   &   $0$         &   $0.032(49)$   &   0.5519 \\
    $J_1$-$\Delta$              &   $0.312(21)$       &   $0$         &   $0$         &   $0.043(47)$   &   0.6113 \\
    $J_1$                       &   $0.326(22)$       &   $0$         &   $0$         &   $0$         &   0.9893 \\
    \bottomrule
    \label{tab:fitparams}
    \end{tabular}
\end{table*}
\begin{figure*}[t]
\includegraphics[width=\linewidth, keepaspectratio]{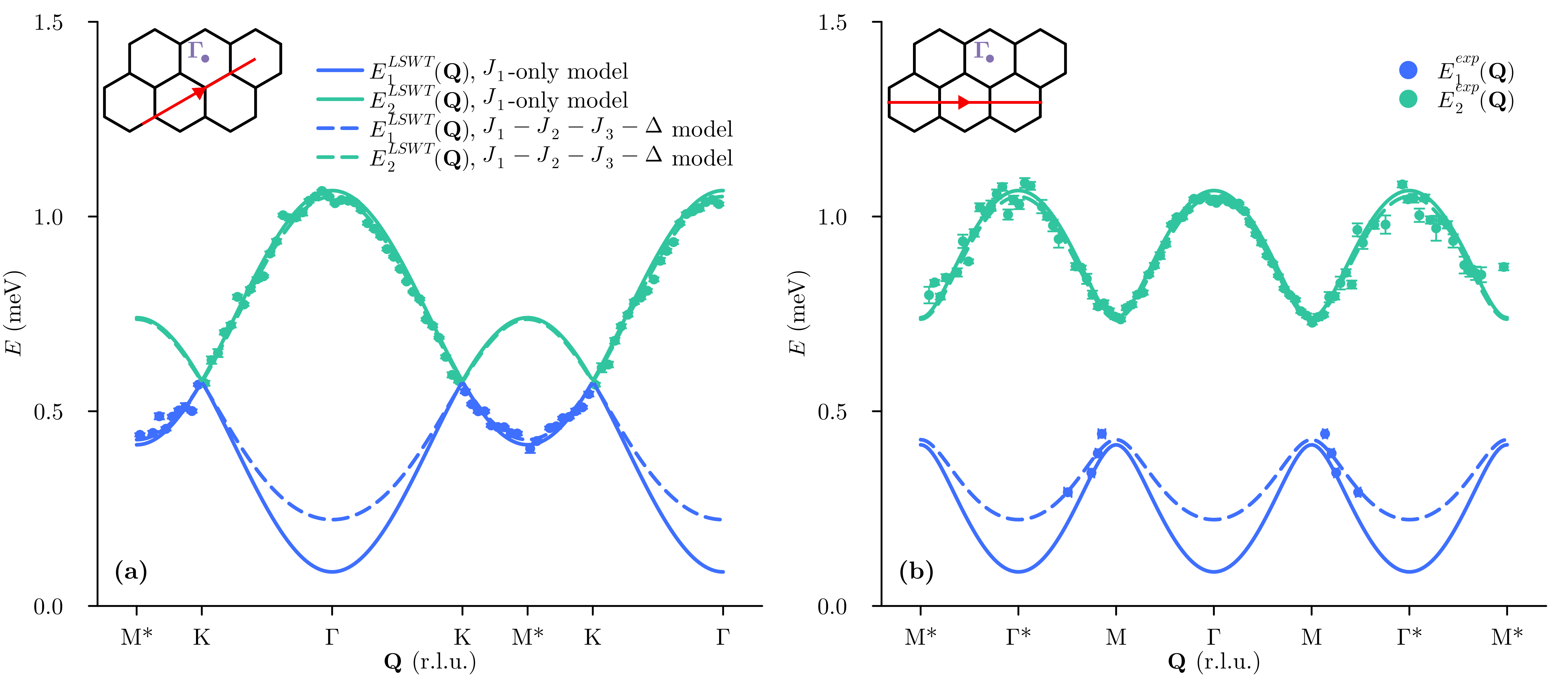}
\caption{{\bf Determination of the spin Hamiltonian of \ce{YbBr3} using LSWT in the field-polarized phase.}
Data points show the magnon energies extracted from CAMEA measurements of the spectrum in the fully field-polarized phase (at $B_{z} = 9.73$ T) for $186$ different $\mathbf{Q}$ vectors along the two high-symmetry reciprocal-lattice directions. Error bars represent an uncertainty of one standard deviation in the fitted peak position. Solid lines show the optimal fit to the magnon dispersion obtained using the nearest-neighbor HHAF Hamiltonian shown in Eq.~(1) of the main text with $J_1 = 0.326$ meV. Dashed lines show the optimal fit obtained using the $J_1$-$J_2$-$J_3$-$\Delta$ model of Eq.~\ref{eq:supham} with the parameters specified in Table \ref{tab:fitparams}. Insets depict the two high-symmetry directions in the reciprocal lattice.}
\label{sup:lswt_disp_fit}
\end{figure*}

The general model we consider for the spin Hamiltonian of \ce{YbBr3} is then the $J_{1}$-$J_{2}$-$J_{3}$-$\Delta$ model
\begin{align}\label{eq:supham}
\begin{split}
    \hat{\mathcal{H}} = & J_{1} \sum_{\langle i,j \rangle}\hat{\mathbf{S}}_{i} \cdot \hat{\mathbf{S}}_{j} + \Delta \hat{S}_{i}^{z} \hat{S}_{j}^{z} \\
                        & + J_{2} \sum_{\langle\langle i,j \rangle\rangle} \hat{\mathbf{S}}_{i} \cdot \hat{\mathbf{S}}_{j}
                          + J_{3}\sum_{\langle\langle\langle i,j \rangle\rangle\rangle} \hat{\mathbf{S}}_{i} \cdot \hat{\mathbf{S}}_{j} \\
                        & - g_{zz} \muB B_{z} \sum_{i} \hat{S}^{z}_{i}.
\end{split}
\end{align}
The interactions $J_{1}$, $J_2$, and $J_{3}$ are Heisenberg superexchange terms between ions in the honeycomb plane on the first-, second-, and third-neighbor sites, which have respective  bond distances of 4.028 $\rm{\AA}$, 6.977 $\rm{\AA}$, and 8.056 $\rm{\AA}$. The parameter $\Delta$ admits an XXZ-type anisotropy on the nearest-neighbor interaction term and is expressed in the form of a departure from Heisenberg physics ($\Delta = 0$), such that $\Delta < 0$ would correspond to XY physics and $\Delta > 0$ to the Ising regime. We neglect the dipolar interactions that were evaluated in Ref.~\cite{Wessler2020} and found to be very small.

The $zz$-component of the $g$-tensor can be determined directly from a measurement of the field-dependence of the Larmor mode, $E_{\mathbf{Q} = \Gamma} = g_{zz}\muB B_{z}$, which is shown in Fig.~1(d) of the main text. Details of the fitting procedure are shown in Fig.~\ref{sup:gzz}, and allow the accurate and independent estimate $g_{zz} = 1.894(26)$. We note here that $g_{zz}$ can be deduced from the CEF parameters, and a calculation based on the parameters reported for \ce{YbBr3} in Ref.~\cite{Wessler2020} yields the value $g_{zz} = 1.6$. Because the extraction of CEF parameters is often an underconstrained problem, the field-dependence of the Larmor mode provides a more precise determination and hence we use the new estimate in our fitting procedure.

We then extract the four remaining parameters in Eq.~(\ref{eq:supham}) by fitting the magnon dispersion calculated for field-polarized phase with LSWT to the magnon dispersion measured on CAMEA at 9.73 T, which is shown in Figs.~1(e-f) of the main text. For this procedure we implemented Eq.~(\ref{eq:supham}) in LSWT using \textsc{Sunny} \cite{sunny} and performed a least-squares minimization. The results of this comparison are shown in Fig.~\ref{sup:lswt_disp_fit} for the two high-symmetry directions in the BZ and the results of our parameter analysis are summarized in Table \ref{tab:fitparams}.

The most striking feature of Fig.~\ref{sup:lswt_disp_fit} is that models with one and four parameters deliver such similar fits to the high-field dispersion. This is reflected in Table \ref{tab:fitparams} by the values of $J_2$, $J_3$, and $\Delta$ all being in the percent range, and smaller than their uncertainties. We remark that these uncertainties, obtained from the covariance matrix at the optimum, are relatively large. Although values of $J_2/J_1$ and $J_3/J_1$ on the order of $5\%$ and $3\%$ respectively cause a noticeable modification of the lower magnon branch in Fig.~\ref{sup:lswt_disp_fit}, our data coverage at these energies is not sufficient to test these possibilities, reflecting the limits to the sensitivity of the fitting procedure. Thus we have no statistically significant evidence for any deviation from a model with $J_2 = J_3 = 0$ when considering the full $\mathbf{Q}$-dependence of the high-field magnon dispersion, and conclude that the optimal and minimal description of our data requires a single, isotropic antiferromagnetic superexchange interaction between nearest-neighbor ions in the honeycomb plane, $J_1 = 0.326 \pm 0.022$ meV.

We remark that it is expected from the compact nature of the electron shells in $f$-electron systems that next-neighbor superexchange interactions are very much smaller than the leading term. Although the near-exact $S = 1/2$ Heisenberg character of a $J = 7/2$ ion such as Yb$^{3+}$ is {\it a priori} somewhat surprising, the separation of crystal-field levels justifies a pseudospin-1/2 description \cite{Wessler2022} and the Heisenberg nature has been explained in the detailed analysis of edge-sharing octahedrally coordinated Yb materials performed in Ref.~\cite{RauGingras2018}. Finally,  we find from a deeper analysis of our MPS data at all fields that the value of $J_1$ we deduce at fields above saturation requires no alteration to describe the spectra at all lower fields, excluding any discernible magnetostriction effects. 

\section{Cylinder MPS calculations}
\label{sup:MPS}

\subsection{Cylinder MPS}

Matrix-Product States (MPS) are a variational Ansatz that can be used to represent the wavefunctions of 1D quantum systems \cite{Schollwoeck2011}. A 2D system can be described by wrapping the lattice onto a cylinder \cite{Stoudenmire2012}. A tensor is assigned to each lattice site and the accuracy of the Ansatz is controlled by the tensor bond dimension, $\chi$, and the cylinder width, $W$, while detailed spatial information is obtained for the direction of the cylinder axis, controlled by its length, $L$. 

We applied cylinder MPS to compute the dynamical spectral function of the isotropic spin-1/2 HHAF in a magnetic field, as defined in Eq.~(1) of the main text, with the energy unit fixed to $J_1 = 0.326$ meV. The cylinder geometry we implement provides a high momentum resolution along the BZ paths $\Gamma$-M-$\Gamma^*$ and $\Gamma^*$-M$^*$ that allow detailed comparison with experiment. The MPS method involves computing the time-dependent real-space spin-spin correlation function
\begin{equation}
C^{\alpha \beta}_\mathbf{r} (\mathbf{x}, t) = \braket{\hat{S}^\alpha_{\mathbf{r} + \mathbf{x}}(t) \hat{S}^\beta_\mathbf{r}(0)},
\end{equation}
in which $\mathbf{r}$ is a site at the center of the cylinder where the initial spin operator is applied, $\mathbf{x}$ is the vector separation in the two-point correlator, and $\alpha,\beta \in \{x, y, z\}$. The U(1) symmetry of the Heisenberg Hamiltonian in a field means that the total magnetization in the field ($z$) direction is conserved under time evolution, which allows us to make our calculations more efficient by working with MPS in a single magnetization sector. Under these circumstances, only the correlation functions with $\alpha\beta \in \{zz, +-, -+\}$ are finite and the others are obtained from standard identities. The calculations were implemented in Python using the package \textsc{TenPy} \cite{Hauschild2018}. 

\begin{figure*}
\includegraphics{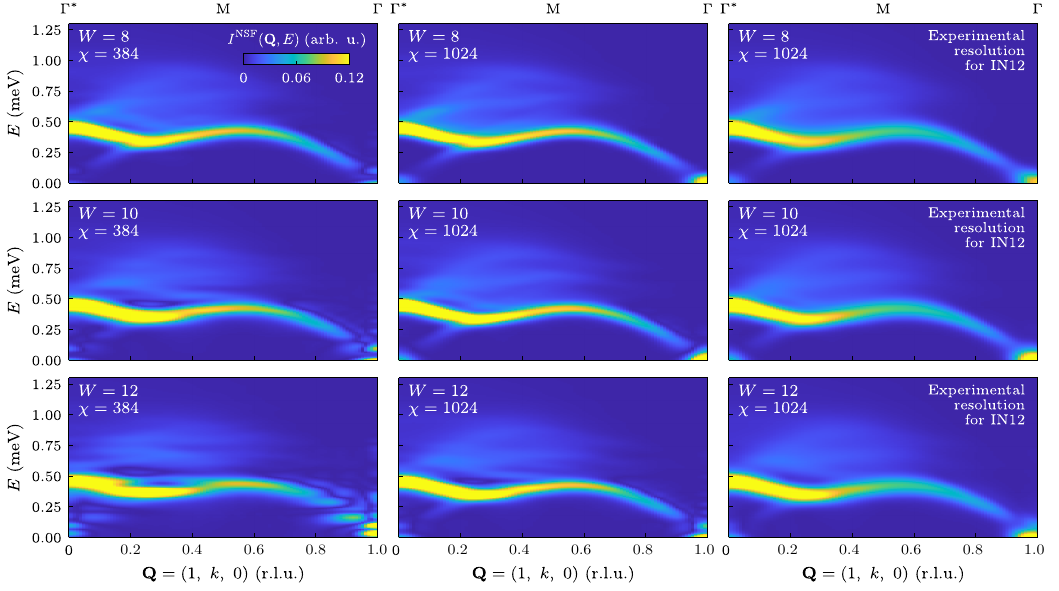}
\caption{{\bf Comparison of MPS spectra computed for different system sizes and bond dimensions.} INS cross-section in the longitudinal channel, $I^{\rm NSF} (\mathbf{Q},E)$, shown along the high-symmetry path $\Gamma^*$-M-$\Gamma$ at a magnetic field corresponding to the experimental value $B = 3.8$~T. The rows show increasing values of the cylinder width, $W$. The left and middle columns compare low and high bond dimensions, $\chi$, with a cautious Gaussian filter applied, while the right column shows the effect of convolution with the experimental resolution function for our highest $\chi$ value. The lower right panel is Fig.~4(b) of the main text.}
\label{LWchi}
\end{figure*}

The dynamical structure factor is obtained from the double Fourier transform
\begin{equation}
\label{eq:cft}
S_{\mathbf{r}}^{\alpha \beta}(\mathbf{Q}, E) = \int_{-\infty}^{\infty} dt \sum_\mathbf{x} e^{i(E t - \mathbf{Q} \cdot \mathbf{x})} C^{\alpha \beta}_\mathbf{r}(\mathbf{x}, t),
\end{equation}
where we retain the subscript $\mathbf{r}$ because the correlation function is computed by time-evolving a ground state that breaks the translational symmetry of the Heisenberg model. This symmetry is restored in the spectral function by averaging over the two different time-evolved states corresponding to the two sites in the central unit cell of the cylinder, thereby reducing the computational cost in comparison with the time-evolution of one spatially symmetrized state. It is standard to compensate for the finite cylinder length and time step by convolving the correlation function with a Gaussian envelope, 
\begin{equation}
\label{mps_convolution}
S^{\alpha \beta}_\mathbf{r}(\mathbf{Q}, E) \! \rightarrow \!\! \sum_{\mathbf{Q'},E'} e^{\big[ -\frac{(E - E')^2}{2\sigma^2_E (E)} - \frac{(\mathbf{Q} - \mathbf{Q'})^2}{2\sigma^2_Q (E)} \big]} S^{\alpha \beta}_\mathbf{r} (\mathbf{Q'}, E'),
\end{equation}
where the sum is over the $\mathbf{Q}$ points in the reciprocal space of the cylinder and over the finite grid of energies evaluated. At very low energies the MPS structure factor is found to contain significant intrinsic noise, which is treated by giving an energy-dependence to the Gaussian width parameters of the form
\begin{align}
\sigma_E (E) & = 0.024~{\rm meV} \big(1 + \tfrac{3}{2} e^{- 4E^2/J_1^2} \big),  \\
\sigma_Q (E) & = 0.016~{\rm \AA^{-1}} \big(1 + \tfrac{3}{2} e^{-(16/3)Q^2a^2} \big).
\end{align}
These are cautious Gaussian filters with standard deviations far below the experimental resolutions in $\mathbf{Q}$ and $E$. In general, $\sigma_E$ and $\sigma_Q$ should remain small to maintain high precision, but should be sufficiently large to avoid cross-contamination of artifacts when performing the experimental resolution convolution, as we explain next. 

\subsection{Resolution-function convolution}

The cross-section measured by INS is subject to a broadening and scaling of intensity compared to the structure factor of the underlying spin system. These effects vary across the $(\mathbf{Q}, E)$ space in a manner depending on the measurement method and the spectrometer. The quantitative agreement between our INS and MPS results shown in the main text was achieved by applying the full experimental resolution function to the MPS data.

\begin{figure*}
\includegraphics{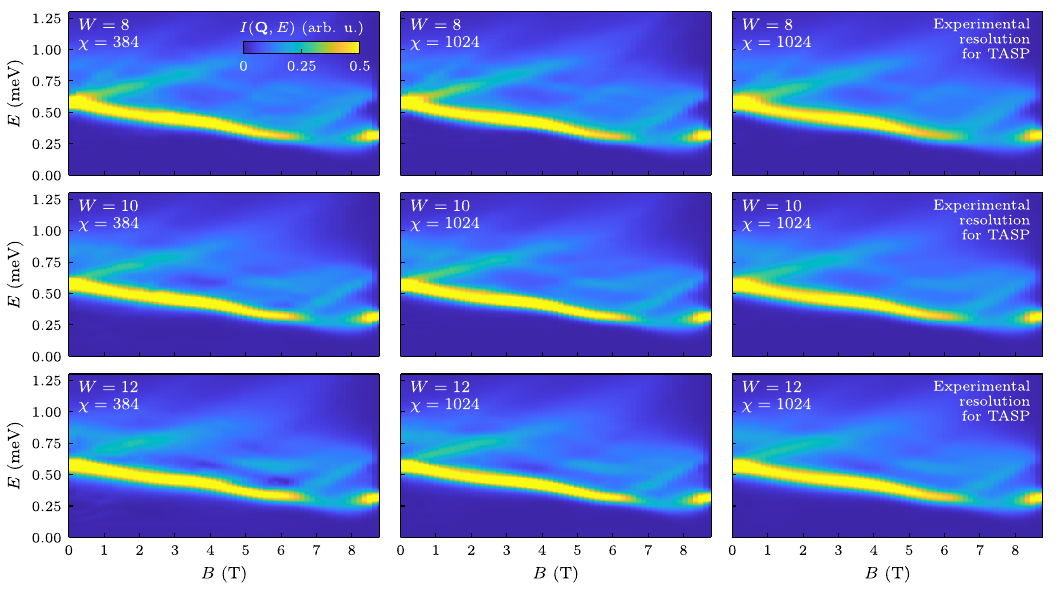}
\caption{{\bf Further comparison of MPS spectra.} Full INS cross-section calculated using different system sizes and bond dimensions at $\bm{Q} = \rm{M^*}$ for all magnetic fields up to saturation. The panel arrangement and data convolution are as in Fig.~\ref{LWchi}. The lower right panel is Fig.~2(b) of the main text.}
\label{LWchi2}
\end{figure*}

We assume that the effective structure factor, $\tilde{S}$, measured by the detector is given by the convolution of the theoretical structure factor, $S$, with a multivariate Gaussian distribution of the form \cite{Popovici1975,Eckold2014} 
\begin{align}
\label{mps_takin}
\tilde{S}(\bm{Q}, E) = R_0 \! \sum_{\bm{Q}', E'} \! S(\bm{Q}' + \bm{Q}, E') \, \rho_0^{-1} \, e^{- \bm{X}^T M \bm{X}} \! ,
\end{align}
where we have suppressed the indices $\alpha, \beta$ because the calculation is the same for all channels. Again $\bm{Q}'$ and $E'$ are summed over the points available from MPS. At each ($\bm{Q}, E$) point, the resolution algorithm should provide distribution parameters $R_0 (\bm{Q},E)$ and $M(\bm{Q},E)$, the latter a 4$\times$4 Gaussian matrix that is expressed as a quadratic form using the difference vector $\bm{X} = (\bm{Q}' - \bm{Q}, E' - E)$. The distribution is normalized by setting $\rho_0(\bm{Q}, E) = \sum_{\bm{Q}', E'} \exp [- \bm{X}^T M \bm{X}]$. In the present work we computed $R_0$ and $M$ using the software tool \textsc{Takin} \cite{Weber2016}.

The fact that $S$ obeys the lattice symmetries allows some additional freedom to shift the momentum transfer ($\bm{Q}$) at which the evaluation is performed. For cylinder MPS calculations, where a limited set of momenta is available, this permits some improvements in the reciprocal-space sampling. Special care is required to correctly rotate $M$ so that it represents the true distribution around $\bm{Q}' + \bm{Q}$. It is crucial to use as input a function $S^{\alpha \beta}_\mathbf{r}(\mathbf{Q}, E)$ that has already been filtered by the prior $Q$ and $E$ convolution specified in Eq.~(\ref{mps_convolution}), as otherwise the non-diagonal resolution matrix $M$ would cause cross-contamination of pre-existing inaccuracies in both coordinates, resulting in uncontrollable artifacts. 

Finally, for comparison with our unpolarized INS measurements, we converted the components of the dynamical structure factor from MPS into a differential cross-section using the expression 
\begin{equation}
\hspace{-8pt} \frac{d^2\sigma}{d\Omega d\omega} \! \propto\ \! |F(\mathbf{Q})|^2 \! \sum_{\alpha,\beta} \! \left( \! \delta_{\alpha\beta} - \frac{Q_{\alpha}Q_{\beta}}{Q^2} \! \right) \!  (g_{\alpha\beta})^2 S_{\alpha\beta}(\mathbf{Q},\omega), \nonumber
\label{Eq:cross-section}
\end{equation}
with $F(\mathbf{Q})$ the magnetic form factor of the Yb$^{3+}$ ion, $(\delta_{\alpha\beta} - Q_{\alpha} Q_{\beta}/Q^2)$ the INS polarization factor, and $g_{\alpha\beta}$ the components of the $g$-tensor deduced from the field-dependence of the spin-precession mode and from measurements of the static magnetic susceptibility \cite{Wessler2020}.

In the main text we showed that the cross-section computed by MPS provides a quantitative account of all the spectral data measured in our INS experiments, meaning for all wave vectors, energies, and applied fields. We stress that the data comparisons are made with a single overall scale factor, $I$, which is independent of {\bf Q}, $E$, and $B$, and changes only when changing the spectrometer. To determine $I$ for each instrument, we performed a single least-squares fit using all the INS intensity data uncontaminated by the elastic signal, for which we used the criteria $E > 0.3$ meV and $Q > 0.22$ $\rm{\AA^{-1}}$, and for all field values simultaneously. To ensure an appropriate comparison of datasets, special care was taken to implement a linear interpolation procedure to evaluate the MPS data at arbitrary values of $\bf{Q}$, $E$, and $B$ while still correctly applying the experimental resolution filter. Using over 2000 experimental data points measured on TASP and over 1100 from IN12, we obtained the respective values $I_{\rm{TASP}} = 1.5368(54)$ and $I_{\rm{IN12}} = 3.128(43)$, from whose uncertainties we establish that the accuracy of a single-parameter fit to all our TASP data is $\sigma_{I,\rm{rel}} = 0.26\%$, and to all our IN12 data is $\sigma_{I,\rm{rel}} = 1.25\%$. 

\subsection{Computational details and physical results}

We computed spectral functions using parameter values up to $W = 12$, $L = 30$, and $\chi = 1024$, and in field steps of size $\Delta B = B_{\rm sat}/30 \equiv 0.293$ T. All correlation functions were evaluated up to a final time $t_\textrm{max} = 90/J_1$, with a time step of $\Delta t = 0.1/J_1$. In Fig.~\ref{LWchi} we use the example of the longitudinal response shown in Fig.~4(b) of the main text to benchmark the cylinder widths, bond dimensions, and time-evolution procedures required to obtain well converged spectral functions. In the left and center columns we apply minimal filter parameters to deduce the intrinsic precision of our MPS calculations. Taking $W = 12$ and $\chi = 384$ as an example of MPS artifacts on the magnon branch around $0.2 < k < 0.4$ and $0.8 < k < 1$, we see that our $\chi = 1024$ spectra are well converged at $W = 8$ and 10, but that weak artifacts may persist at $W = 12$. Taking the object of our study as being to assure that the physical features we observe are robust, rather than to achieve absolute quantitative precision beyond the level of the effective energy and momentum resolutions of our INS experiments, we include the appropriate resolution functions in the right column of Fig.~\ref{LWchi}. At this level, convergence of all the observed features is achieved using our higher $W$ and $\chi$ values, as we verify over the full field range for the example of the M$^*$ point in Fig.~\ref{LWchi2}. All of the results shown in Figs.~2-4 of the main text were obtained using $W = 12$ and $L = 30$, with $\chi$ set to 1024, with the exception of the Z$^*$ point in Fig.~3(d), where $W = 10$. 

Finally, for the identification of the roton-like excitation it was important to look at the separate transverse and longitudinal spectral functions computed by MPS, respectively $S_{xx}(\mathbf{Q},E)$ and $S_{zz}(\mathbf{Q},E)$, and in Fig.~4 of the main text we showed only the longitudinal response for comparison with polarized INS measurements. To illustrate how the spectral features visible in the transverse and longitudinal channels change in a complementary manner across the BZ, we used the MPS $S_{xx}(\mathbf{Q},E)$ and $S_{zz}(\mathbf{Q},E)$ to compute the experimental observables of polarized INS, $I^{\rm{SF}}(\mathbf{Q},E)$ (spin-flip) and $I^{\rm{NSF}}(\mathbf{Q},E)$ (non-spin-flip). In Fig.~\ref{sup:in12_5T} we show the transverse (SF) response [Fig.~\ref{sup:in12_5T}(d)] corresponding to the longitudinal (NSF) response [Fig.~\ref{sup:in12_5T}(e)] computed by MPS for a field of 5 T, which was shown in Fig.~4(c) of the main text. Figure \ref{sup:in12_5T}(f) shows the sum of the two channels, illustrating that the unpolarized response also reflects the unconventional departure of spectral weight from the LSWT magnon branches that otherwise provide a qualitative guide to the modes in the inner BZ. Figures \ref{sup:in12_5T}(a-c) show the INS measurements described quantitatively by these MPS results. 


\begin{thebibliography}{68}%
\makeatletter
\providecommand \@ifxundefined [1]{%
 \@ifx{#1\undefined}
}%
\providecommand \@ifnum [1]{%
 \ifnum #1\expandafter \@firstoftwo
 \else \expandafter \@secondoftwo
 \fi
}%
\providecommand \@ifx [1]{%
 \ifx #1\expandafter \@firstoftwo
 \else \expandafter \@secondoftwo
 \fi
}%
\providecommand \natexlab [1]{#1}%
\providecommand \enquote  [1]{``#1''}%
\providecommand \bibnamefont  [1]{#1}%
\providecommand \bibfnamefont [1]{#1}%
\providecommand \citenamefont [1]{#1}%
\providecommand \href@noop [0]{\@secondoftwo}%
\providecommand \href [0]{\begingroup \@sanitize@url \@href}%
\providecommand \@href[1]{\@@startlink{#1}\@@href}%
\providecommand \@@href[1]{\endgroup#1\@@endlink}%
\providecommand \@sanitize@url [0]{\catcode `\\12\catcode `\$12\catcode
  `\&12\catcode `\#12\catcode `\^12\catcode `\_12\catcode `\%12\relax}%
\providecommand \@@startlink[1]{}%
\providecommand \@@endlink[0]{}%
\providecommand \url  [0]{\begingroup\@sanitize@url \@url }%
\providecommand \@url [1]{\endgroup\@href {#1}{\urlprefix }}%
\providecommand \urlprefix  [0]{URL }%
\providecommand \Eprint [0]{\href }%
\providecommand \doibase [0]{https://doi.org/}%
\providecommand \selectlanguage [0]{\@gobble}%
\providecommand \bibinfo  [0]{\@secondoftwo}%
\providecommand \bibfield  [0]{\@secondoftwo}%
\providecommand \translation [1]{[#1]}%
\providecommand \BibitemOpen [0]{}%
\providecommand \bibitemStop [0]{}%
\providecommand \bibitemNoStop [0]{.\EOS\space}%
\providecommand \EOS [0]{\spacefactor3000\relax}%
\providecommand \BibitemShut  [1]{\csname bibitem#1\endcsname}%
\let\auto@bib@innerbib\@empty
%</preamble>
\bibitem [{\citenamefont {Savary}\ and\ \citenamefont
  {Balents}(2016)}]{Savary2016}%
  \BibitemOpen
  \bibfield  {author} {\bibinfo {author} {\bibfnamefont {L.}~\bibnamefont
  {Savary}}\ and\ \bibinfo {author} {\bibfnamefont {L.}~\bibnamefont
  {Balents}},\ }\bibfield  {title} {\bibinfo {title} {{Quantum spin liquids: a
  review}},\ }\href {https://doi.org/10.1088/0034-4885/80/1/016502} {\bibfield
  {journal} {\bibinfo  {journal} {Rep. Prog. Phys.}\ }\textbf {\bibinfo
  {volume} {80}},\ \bibinfo {pages} {016502} (\bibinfo {year}
  {2016})}\BibitemShut {NoStop}%
\bibitem [{\citenamefont {Broholm}\ \emph {et~al.}(2020)\citenamefont
  {Broholm}, \citenamefont {Cava}, \citenamefont {Kivelson}, \citenamefont
  {Nocera}, \citenamefont {Norman},\ and\ \citenamefont
  {Senthil}}]{Broholm2020}%
  \BibitemOpen
  \bibfield  {author} {\bibinfo {author} {\bibfnamefont {C.}~\bibnamefont
  {Broholm}}, \bibinfo {author} {\bibfnamefont {R.~J.}\ \bibnamefont {Cava}},
  \bibinfo {author} {\bibfnamefont {S.~A.}\ \bibnamefont {Kivelson}}, \bibinfo
  {author} {\bibfnamefont {D.~G.}\ \bibnamefont {Nocera}}, \bibinfo {author}
  {\bibfnamefont {M.~R.}\ \bibnamefont {Norman}},\ and\ \bibinfo {author}
  {\bibfnamefont {T.}~\bibnamefont {Senthil}},\ }\bibfield  {title} {\bibinfo
  {title} {{Quantum spin liquids}},\ }\href
  {https://doi.org/10.1126/science.aay0668} {\bibfield  {journal} {\bibinfo
  {journal} {Science}\ }\textbf {\bibinfo {volume} {367}},\ \bibinfo {pages}
  {eaay0668} (\bibinfo {year} {2020})}\BibitemShut {NoStop}%
\bibitem [{\citenamefont {Zhitomirsky}\ and\ \citenamefont
  {Chernyshev}(1999)}]{zc1999}%
  \BibitemOpen
  \bibfield  {author} {\bibinfo {author} {\bibfnamefont {M.~E.}\ \bibnamefont
  {Zhitomirsky}}\ and\ \bibinfo {author} {\bibfnamefont {A.~L.}\ \bibnamefont
  {Chernyshev}},\ }\bibfield  {title} {\bibinfo {title} {{Instability of
  Antiferromagnetic Magnons in Strong fields}},\ }\href
  {https://doi.org/10.1103/PhysRevLett.82.4536} {\bibfield  {journal} {\bibinfo
   {journal} {Phys. Rev. Lett.}\ }\textbf {\bibinfo {volume} {82}},\ \bibinfo
  {pages} {4536} (\bibinfo {year} {1999})}\BibitemShut {NoStop}%
\bibitem [{\citenamefont {L\"uscher}\ and\ \citenamefont
  {L\"auchli}(2009)}]{Luescher2009}%
  \BibitemOpen
  \bibfield  {author} {\bibinfo {author} {\bibfnamefont {A.}~\bibnamefont
  {L\"uscher}}\ and\ \bibinfo {author} {\bibfnamefont {A.~M.}\ \bibnamefont
  {L\"auchli}},\ }\bibfield  {title} {\bibinfo {title} {{Exact diagonalization
  study of the antiferromagnetic spin-1/2 Heisenberg model on the square
  lattice in a magnetic field}},\ }\href
  {https://doi.org/10.1103/PhysRevB.79.195102} {\bibfield  {journal} {\bibinfo
  {journal} {Phys. Rev. B}\ }\textbf {\bibinfo {volume} {79}},\ \bibinfo
  {pages} {195102} (\bibinfo {year} {2009})}\BibitemShut {NoStop}%
\bibitem [{\citenamefont {Mourigal}\ \emph {et~al.}(2010)\citenamefont
  {Mourigal}, \citenamefont {Zhitomirsky},\ and\ \citenamefont
  {Chernyshev}}]{mzc2010}%
  \BibitemOpen
  \bibfield  {author} {\bibinfo {author} {\bibfnamefont {M.}~\bibnamefont
  {Mourigal}}, \bibinfo {author} {\bibfnamefont {M.~E.}\ \bibnamefont
  {Zhitomirsky}},\ and\ \bibinfo {author} {\bibfnamefont {A.~L.}\ \bibnamefont
  {Chernyshev}},\ }\bibfield  {title} {\bibinfo {title} {Field-induced decay
  dynamics in square-lattice antiferromagnets},\ }\href
  {https://doi.org/10.1103/PhysRevB.82.144402} {\bibfield  {journal} {\bibinfo
  {journal} {Phys. Rev. B}\ }\textbf {\bibinfo {volume} {82}},\ \bibinfo
  {pages} {144402} (\bibinfo {year} {2010})}\BibitemShut {NoStop}%
\bibitem [{\citenamefont {Fuhrman}\ \emph {et~al.}(2012)\citenamefont
  {Fuhrman}, \citenamefont {Mourigal}, \citenamefont {Zhitomirsky},\ and\
  \citenamefont {Chernyshev}}]{fmzc2012}%
  \BibitemOpen
  \bibfield  {author} {\bibinfo {author} {\bibfnamefont {W.~T.}\ \bibnamefont
  {Fuhrman}}, \bibinfo {author} {\bibfnamefont {M.}~\bibnamefont {Mourigal}},
  \bibinfo {author} {\bibfnamefont {M.~E.}\ \bibnamefont {Zhitomirsky}},\ and\
  \bibinfo {author} {\bibfnamefont {A.~L.}\ \bibnamefont {Chernyshev}},\
  }\bibfield  {title} {\bibinfo {title} {Dynamical structure factor of
  quasi-two-dimensional antiferromagnet in high fields},\ }\href
  {https://doi.org/10.1103/PhysRevB.85.184405} {\bibfield  {journal} {\bibinfo
  {journal} {Phys. Rev. B}\ }\textbf {\bibinfo {volume} {85}},\ \bibinfo
  {pages} {184405} (\bibinfo {year} {2012})}\BibitemShut {NoStop}%
\bibitem [{\citenamefont {Zhitomirsky}\ and\ \citenamefont
  {Chernyshev}(2013)}]{zc2013}%
  \BibitemOpen
  \bibfield  {author} {\bibinfo {author} {\bibfnamefont {M.~E.}\ \bibnamefont
  {Zhitomirsky}}\ and\ \bibinfo {author} {\bibfnamefont {A.~L.}\ \bibnamefont
  {Chernyshev}},\ }\bibfield  {title} {\bibinfo {title} {{Colloquium:
  Spontaneous magnon decays}},\ }\href
  {https://doi.org/10.1103/RevModPhys.85.219} {\bibfield  {journal} {\bibinfo
  {journal} {Rev. Mod. Phys.}\ }\textbf {\bibinfo {volume} {85}},\ \bibinfo
  {pages} {219} (\bibinfo {year} {2013})}\BibitemShut {NoStop}%
\bibitem [{\citenamefont {Headings}\ \emph {et~al.}(2010)\citenamefont
  {Headings}, \citenamefont {Hayden}, \citenamefont {Coldea},\ and\
  \citenamefont {Perring}}]{Headings2010}%
  \BibitemOpen
  \bibfield  {author} {\bibinfo {author} {\bibfnamefont {N.~S.}\ \bibnamefont
  {Headings}}, \bibinfo {author} {\bibfnamefont {S.~M.}\ \bibnamefont
  {Hayden}}, \bibinfo {author} {\bibfnamefont {R.}~\bibnamefont {Coldea}},\
  and\ \bibinfo {author} {\bibfnamefont {T.~G.}\ \bibnamefont {Perring}},\
  }\bibfield  {title} {\bibinfo {title} {{Anomalous High-Energy Spin
  Excitations in the High-$T_c$ Superconductor-Parent Antiferromagnet
  La$_2$CuO$_4$}},\ }\href {https://doi.org/10.1103/PhysRevLett.105.247001}
  {\bibfield  {journal} {\bibinfo  {journal} {Phys. Rev. Lett.}\ }\textbf
  {\bibinfo {volume} {105}},\ \bibinfo {pages} {247001} (\bibinfo {year}
  {2010})}\BibitemShut {NoStop}%
\bibitem [{\citenamefont {Plumb}\ \emph {et~al.}(2014)\citenamefont {Plumb},
  \citenamefont {Savici}, \citenamefont {Granroth}, \citenamefont {Chou},\ and\
  \citenamefont {Kim}}]{Plumb2014}%
  \BibitemOpen
  \bibfield  {author} {\bibinfo {author} {\bibfnamefont {K.~W.}\ \bibnamefont
  {Plumb}}, \bibinfo {author} {\bibfnamefont {A.~T.}\ \bibnamefont {Savici}},
  \bibinfo {author} {\bibfnamefont {G.~E.}\ \bibnamefont {Granroth}}, \bibinfo
  {author} {\bibfnamefont {F.~C.}\ \bibnamefont {Chou}},\ and\ \bibinfo
  {author} {\bibfnamefont {Y.-J.}\ \bibnamefont {Kim}},\ }\bibfield  {title}
  {\bibinfo {title} {{High-energy continuum of magnetic excitations in the
  two-dimensional quantum antiferromagnet Sr$_2$CuO$_2$Cl$_2$}},\ }\href
  {https://doi.org/10.1103/PhysRevB.89.180410} {\bibfield  {journal} {\bibinfo
  {journal} {Phys. Rev. B}\ }\textbf {\bibinfo {volume} {89}},\ \bibinfo
  {pages} {180410(R)} (\bibinfo {year} {2014})}\BibitemShut {NoStop}%
\bibitem [{\citenamefont {Dalla~Piazza}\ \emph {et~al.}(2014)\citenamefont
  {Dalla~Piazza}, \citenamefont {Mourigal}, \citenamefont {Christensen},
  \citenamefont {Nilsen}, \citenamefont {Tregenna-Piggott}, \citenamefont
  {Perring}, \citenamefont {Enderle}, \citenamefont {McMorrow}, \citenamefont
  {Ivanov},\ and\ \citenamefont {R{\o}nnow}}]{DallaPiazza2014}%
  \BibitemOpen
  \bibfield  {author} {\bibinfo {author} {\bibfnamefont {B.}~\bibnamefont
  {Dalla~Piazza}}, \bibinfo {author} {\bibfnamefont {M.}~\bibnamefont
  {Mourigal}}, \bibinfo {author} {\bibfnamefont {N.~B.}\ \bibnamefont
  {Christensen}}, \bibinfo {author} {\bibfnamefont {G.~J.}\ \bibnamefont
  {Nilsen}}, \bibinfo {author} {\bibfnamefont {P.}~\bibnamefont
  {Tregenna-Piggott}}, \bibinfo {author} {\bibfnamefont {T.~G.}\ \bibnamefont
  {Perring}}, \bibinfo {author} {\bibfnamefont {M.}~\bibnamefont {Enderle}},
  \bibinfo {author} {\bibfnamefont {D.~F.}\ \bibnamefont {McMorrow}}, \bibinfo
  {author} {\bibfnamefont {D.~A.}\ \bibnamefont {Ivanov}},\ and\ \bibinfo
  {author} {\bibfnamefont {H.~M.}\ \bibnamefont {R{\o}nnow}},\ }\bibfield
  {title} {\bibinfo {title} {Fractional excitations in the square-lattice
  quantum antiferromagnet},\ }\href {http://dx.doi.org/10.1038/nphys3172}
  {\bibfield  {journal} {\bibinfo  {journal} {Nat. Phys.}\ }\textbf {\bibinfo
  {volume} {11}},\ \bibinfo {pages} {62} (\bibinfo {year} {2014})}\BibitemShut
  {NoStop}%
\bibitem [{\citenamefont {Powalski}\ \emph {et~al.}(2015)\citenamefont
  {Powalski}, \citenamefont {Uhrig},\ and\ \citenamefont
  {Schmidt}}]{Powalski2016}%
  \BibitemOpen
  \bibfield  {author} {\bibinfo {author} {\bibfnamefont {M.}~\bibnamefont
  {Powalski}}, \bibinfo {author} {\bibfnamefont {G.~S.}\ \bibnamefont
  {Uhrig}},\ and\ \bibinfo {author} {\bibfnamefont {K.~P.}\ \bibnamefont
  {Schmidt}},\ }\bibfield  {title} {\bibinfo {title} {{Roton Minimum as a
  Fingerprint of Magnon-Higgs Scattering in Ordered Quantum
  Antiferromagnets}},\ }\href {https://doi.org/10.1103/PhysRevLett.115.207202}
  {\bibfield  {journal} {\bibinfo  {journal} {Phys. Rev. Lett.}\ }\textbf
  {\bibinfo {volume} {115}},\ \bibinfo {pages} {207202} (\bibinfo {year}
  {2015})}\BibitemShut {NoStop}%
\bibitem [{\citenamefont {Powalski}\ \emph {et~al.}(2018)\citenamefont
  {Powalski}, \citenamefont {Schmidt},\ and\ \citenamefont
  {Uhrig}}]{Powalski2018}%
  \BibitemOpen
  \bibfield  {author} {\bibinfo {author} {\bibfnamefont {M.}~\bibnamefont
  {Powalski}}, \bibinfo {author} {\bibfnamefont {K.~P.}\ \bibnamefont
  {Schmidt}},\ and\ \bibinfo {author} {\bibfnamefont {G.~S.}\ \bibnamefont
  {Uhrig}},\ }\bibfield  {title} {\bibinfo {title} {{Mutually attracting spin
  waves in the square-lattice quantum antiferromagnet}},\ }\href
  {https://doi.org/10.21468/SciPostPhys.4.1.001} {\bibfield  {journal}
  {\bibinfo  {journal} {SciPost Phys.}\ }\textbf {\bibinfo {volume} {4}},\
  \bibinfo {pages} {001} (\bibinfo {year} {2018})}\BibitemShut {NoStop}%
\bibitem [{\citenamefont {Auerbach}\ and\ \citenamefont
  {Arovas}(1988)}]{AA1988}%
  \BibitemOpen
  \bibfield  {author} {\bibinfo {author} {\bibfnamefont {A.}~\bibnamefont
  {Auerbach}}\ and\ \bibinfo {author} {\bibfnamefont {D.~P.}\ \bibnamefont
  {Arovas}},\ }\bibfield  {title} {\bibinfo {title} {{Spin Dynamics in the
  Square-Lattice Antiferromagnet}},\ }\href
  {https://doi.org/10.1103/PhysRevLett.61.617} {\bibfield  {journal} {\bibinfo
  {journal} {Phys. Rev. Lett.}\ }\textbf {\bibinfo {volume} {61}},\ \bibinfo
  {pages} {617} (\bibinfo {year} {1988})}\BibitemShut {NoStop}%
\bibitem [{\citenamefont {Hsu}(1990)}]{Hsu1990}%
  \BibitemOpen
  \bibfield  {author} {\bibinfo {author} {\bibfnamefont {T.~C.}\ \bibnamefont
  {Hsu}},\ }\bibfield  {title} {\bibinfo {title} {{Spin waves in the flux-phase
  description of the $S = 1/2$ Heisenberg antiferromagnet}},\ }\href
  {https://doi.org/10.1103/PhysRevB.41.11379} {\bibfield  {journal} {\bibinfo
  {journal} {Phys. Rev. B}\ }\textbf {\bibinfo {volume} {41}},\ \bibinfo
  {pages} {11379} (\bibinfo {year} {1990})}\BibitemShut {NoStop}%
\bibitem [{\citenamefont {Ho}\ \emph {et~al.}(2001)\citenamefont {Ho},
  \citenamefont {Muthukumar}, \citenamefont {Ogata},\ and\ \citenamefont
  {Anderson}}]{Ho2001}%
  \BibitemOpen
  \bibfield  {author} {\bibinfo {author} {\bibfnamefont {C.-M.}\ \bibnamefont
  {Ho}}, \bibinfo {author} {\bibfnamefont {V.~N.}\ \bibnamefont {Muthukumar}},
  \bibinfo {author} {\bibfnamefont {M.}~\bibnamefont {Ogata}},\ and\ \bibinfo
  {author} {\bibfnamefont {P.~W.}\ \bibnamefont {Anderson}},\ }\bibfield
  {title} {\bibinfo {title} {{Nature of Spin Excitations in Two-Dimensional
  Mott Insulators: Undoped Cuprates and Other Materials}},\ }\href
  {https://doi.org/10.1103/PhysRevLett.86.1626} {\bibfield  {journal} {\bibinfo
   {journal} {Phys. Rev. Lett.}\ }\textbf {\bibinfo {volume} {86}},\ \bibinfo
  {pages} {1626} (\bibinfo {year} {2001})}\BibitemShut {NoStop}%
\bibitem [{\citenamefont {Shao}\ \emph {et~al.}(2017)\citenamefont {Shao},
  \citenamefont {Qin}, \citenamefont {Capponi}, \citenamefont {Chesi},
  \citenamefont {Meng},\ and\ \citenamefont {Sandvik}}]{Shao2017}%
  \BibitemOpen
  \bibfield  {author} {\bibinfo {author} {\bibfnamefont {H.}~\bibnamefont
  {Shao}}, \bibinfo {author} {\bibfnamefont {Y.~Q.}\ \bibnamefont {Qin}},
  \bibinfo {author} {\bibfnamefont {S.}~\bibnamefont {Capponi}}, \bibinfo
  {author} {\bibfnamefont {S.}~\bibnamefont {Chesi}}, \bibinfo {author}
  {\bibfnamefont {Z.~Y.}\ \bibnamefont {Meng}},\ and\ \bibinfo {author}
  {\bibfnamefont {A.~W.}\ \bibnamefont {Sandvik}},\ }\bibfield  {title}
  {\bibinfo {title} {{Nearly Deconfined Spinon Excitations in the
  Square-Lattice Spin-1/2 Heisenberg Antiferromagnet}},\ }\href
  {https://doi.org/10.1103/PhysRevX.7.041072} {\bibfield  {journal} {\bibinfo
  {journal} {Phys. Rev. X}\ }\textbf {\bibinfo {volume} {7}},\ \bibinfo {pages}
  {041072} (\bibinfo {year} {2017})}\BibitemShut {NoStop}%
\bibitem [{\citenamefont {Yu}\ \emph {et~al.}(2018)\citenamefont {Yu},
  \citenamefont {Wang}, \citenamefont {Dong}, \citenamefont {Yao},\ and\
  \citenamefont {Li}}]{Yu2018}%
  \BibitemOpen
  \bibfield  {author} {\bibinfo {author} {\bibfnamefont {S.-L.}\ \bibnamefont
  {Yu}}, \bibinfo {author} {\bibfnamefont {W.}~\bibnamefont {Wang}}, \bibinfo
  {author} {\bibfnamefont {Z.-Y.}\ \bibnamefont {Dong}}, \bibinfo {author}
  {\bibfnamefont {Z.-J.}\ \bibnamefont {Yao}},\ and\ \bibinfo {author}
  {\bibfnamefont {J.-X.}\ \bibnamefont {Li}},\ }\bibfield  {title} {\bibinfo
  {title} {Deconfinement of spinons in frustrated spin systems: Spectral
  perspective},\ }\href {https://doi.org/10.1103/PhysRevB.98.134410} {\bibfield
   {journal} {\bibinfo  {journal} {Phys. Rev. B}\ }\textbf {\bibinfo {volume}
  {98}},\ \bibinfo {pages} {134410} (\bibinfo {year} {2018})}\BibitemShut
  {NoStop}%
\bibitem [{\citenamefont {Zhang}\ \emph {et~al.}(2022)\citenamefont {Zhang},
  \citenamefont {Ghioldi}, \citenamefont {Manuel}, \citenamefont {Trumper},\
  and\ \citenamefont {Batista}}]{Zhang2022}%
  \BibitemOpen
  \bibfield  {author} {\bibinfo {author} {\bibfnamefont {S.-S.}\ \bibnamefont
  {Zhang}}, \bibinfo {author} {\bibfnamefont {E.~A.}\ \bibnamefont {Ghioldi}},
  \bibinfo {author} {\bibfnamefont {L.~O.}\ \bibnamefont {Manuel}}, \bibinfo
  {author} {\bibfnamefont {A.~E.}\ \bibnamefont {Trumper}},\ and\ \bibinfo
  {author} {\bibfnamefont {C.~D.}\ \bibnamefont {Batista}},\ }\bibfield
  {title} {\bibinfo {title} {{Schwinger boson theory of ordered magnets}},\
  }\href {https://doi.org/10.1103/PhysRevB.105.224404} {\bibfield  {journal}
  {\bibinfo  {journal} {Phys. Rev. B}\ }\textbf {\bibinfo {volume} {105}},\
  \bibinfo {pages} {224404} (\bibinfo {year} {2022})}\BibitemShut {NoStop}%
\bibitem [{\citenamefont {Fouet}\ \emph {et~al.}(2001)\citenamefont {Fouet},
  \citenamefont {Sindzingre},\ and\ \citenamefont {Lhuillier}}]{Fouet2001}%
  \BibitemOpen
  \bibfield  {author} {\bibinfo {author} {\bibfnamefont {J.~B.}\ \bibnamefont
  {Fouet}}, \bibinfo {author} {\bibfnamefont {P.}~\bibnamefont {Sindzingre}},\
  and\ \bibinfo {author} {\bibfnamefont {C.}~\bibnamefont {Lhuillier}},\
  }\bibfield  {title} {\bibinfo {title} {{An investigation of the quantum
  $J_1$-$J_2$-$J_3$ model on the honeycomb lattice}},\ }\href
  {https://doi.org/10.1007/s100510170273} {\bibfield  {journal} {\bibinfo
  {journal} {Eur. Phys. J. B}\ }\textbf {\bibinfo {volume} {20}},\ \bibinfo
  {pages} {241} (\bibinfo {year} {2001})}\BibitemShut {NoStop}%
\bibitem [{\citenamefont {Mulder}\ \emph {et~al.}(2010)\citenamefont {Mulder},
  \citenamefont {Ganesh}, \citenamefont {Capriotti},\ and\ \citenamefont
  {Paramekanti}}]{Mulder2010}%
  \BibitemOpen
  \bibfield  {author} {\bibinfo {author} {\bibfnamefont {A.}~\bibnamefont
  {Mulder}}, \bibinfo {author} {\bibfnamefont {R.}~\bibnamefont {Ganesh}},
  \bibinfo {author} {\bibfnamefont {L.}~\bibnamefont {Capriotti}},\ and\
  \bibinfo {author} {\bibfnamefont {A.}~\bibnamefont {Paramekanti}},\
  }\bibfield  {title} {\bibinfo {title} {{Spiral order by disorder and lattice
  nematic order in a frustrated Heisenberg antiferromagnet on the honeycomb
  lattice}},\ }\href {https://doi.org/10.1103/PhysRevB.81.214419} {\bibfield
  {journal} {\bibinfo  {journal} {Phys. Rev. B}\ }\textbf {\bibinfo {volume}
  {81}},\ \bibinfo {pages} {214419} (\bibinfo {year} {2010})}\BibitemShut
  {NoStop}%
\bibitem [{\citenamefont {Oitmaa}\ and\ \citenamefont
  {Singh}(2011)}]{Oitmaa2011}%
  \BibitemOpen
  \bibfield  {author} {\bibinfo {author} {\bibfnamefont {J.}~\bibnamefont
  {Oitmaa}}\ and\ \bibinfo {author} {\bibfnamefont {R.~R.~P.}\ \bibnamefont
  {Singh}},\ }\bibfield  {title} {\bibinfo {title} {{Phase diagram of the
  $J_1$-$J_2$-$J_3$ Heisenberg model on the honeycomb lattice: A series
  expansion study}},\ }\href {https://doi.org/10.1103/PhysRevB.84.094424}
  {\bibfield  {journal} {\bibinfo  {journal} {Phys. Rev. B}\ }\textbf {\bibinfo
  {volume} {84}},\ \bibinfo {pages} {094424} (\bibinfo {year}
  {2011})}\BibitemShut {NoStop}%
\bibitem [{\citenamefont {Farnell}\ \emph {et~al.}(2011)\citenamefont
  {Farnell}, \citenamefont {Bishop}, \citenamefont {Li}, \citenamefont
  {Richter},\ and\ \citenamefont {Campbell}}]{Farnell2011}%
  \BibitemOpen
  \bibfield  {author} {\bibinfo {author} {\bibfnamefont {D.~J.~J.}\
  \bibnamefont {Farnell}}, \bibinfo {author} {\bibfnamefont {R.~F.}\
  \bibnamefont {Bishop}}, \bibinfo {author} {\bibfnamefont {P.~H.~Y.}\
  \bibnamefont {Li}}, \bibinfo {author} {\bibfnamefont {J.}~\bibnamefont
  {Richter}},\ and\ \bibinfo {author} {\bibfnamefont {C.~E.}\ \bibnamefont
  {Campbell}},\ }\bibfield  {title} {\bibinfo {title} {{Frustrated Heisenberg
  antiferromagnet on the honeycomb lattice: A candidate for deconfined quantum
  criticality}},\ }\href {https://doi.org/10.1103/PhysRevB.84.012403}
  {\bibfield  {journal} {\bibinfo  {journal} {Phys. Rev. B}\ }\textbf {\bibinfo
  {volume} {84}},\ \bibinfo {pages} {012403} (\bibinfo {year}
  {2011})}\BibitemShut {NoStop}%
\bibitem [{\citenamefont {Albuquerque}\ \emph {et~al.}(2011)\citenamefont
  {Albuquerque}, \citenamefont {Schwandt}, \citenamefont {Het\'enyi},
  \citenamefont {Capponi}, \citenamefont {Mambrini},\ and\ \citenamefont
  {L\"auchli}}]{Albuquerque2011}%
  \BibitemOpen
  \bibfield  {author} {\bibinfo {author} {\bibfnamefont {A.~F.}\ \bibnamefont
  {Albuquerque}}, \bibinfo {author} {\bibfnamefont {D.}~\bibnamefont
  {Schwandt}}, \bibinfo {author} {\bibfnamefont {B.}~\bibnamefont {Het\'enyi}},
  \bibinfo {author} {\bibfnamefont {S.}~\bibnamefont {Capponi}}, \bibinfo
  {author} {\bibfnamefont {M.}~\bibnamefont {Mambrini}},\ and\ \bibinfo
  {author} {\bibfnamefont {A.~M.}\ \bibnamefont {L\"auchli}},\ }\bibfield
  {title} {\bibinfo {title} {{Phase diagram of a frustrated quantum
  antiferromagnet on the honeycomb lattice: Magnetic order versus valence-bond
  crystal formation}},\ }\href {https://doi.org/10.1103/PhysRevB.84.024406}
  {\bibfield  {journal} {\bibinfo  {journal} {Phys. Rev. B}\ }\textbf {\bibinfo
  {volume} {84}},\ \bibinfo {pages} {024406} (\bibinfo {year}
  {2011})}\BibitemShut {NoStop}%
\bibitem [{\citenamefont {Ganesh}\ \emph {et~al.}(2013)\citenamefont {Ganesh},
  \citenamefont {van~den Brink},\ and\ \citenamefont {Nishimoto}}]{Ganesh2013}%
  \BibitemOpen
  \bibfield  {author} {\bibinfo {author} {\bibfnamefont {R.}~\bibnamefont
  {Ganesh}}, \bibinfo {author} {\bibfnamefont {J.}~\bibnamefont {van~den
  Brink}},\ and\ \bibinfo {author} {\bibfnamefont {S.}~\bibnamefont
  {Nishimoto}},\ }\bibfield  {title} {\bibinfo {title} {{Deconfined Criticality
  in the Frustrated Heisenberg Honeycomb Antiferromagnet}},\ }\href
  {https://doi.org/10.1103/PhysRevLett.110.127203} {\bibfield  {journal}
  {\bibinfo  {journal} {Phys. Rev. Lett.}\ }\textbf {\bibinfo {volume} {110}},\
  \bibinfo {pages} {127203} (\bibinfo {year} {2013})}\BibitemShut {NoStop}%
\bibitem [{\citenamefont {Zhu}\ \emph {et~al.}(2013)\citenamefont {Zhu},
  \citenamefont {Huse},\ and\ \citenamefont {White}}]{Zhu2013}%
  \BibitemOpen
  \bibfield  {author} {\bibinfo {author} {\bibfnamefont {Z.}~\bibnamefont
  {Zhu}}, \bibinfo {author} {\bibfnamefont {D.~A.}\ \bibnamefont {Huse}},\ and\
  \bibinfo {author} {\bibfnamefont {S.~R.}\ \bibnamefont {White}},\ }\bibfield
  {title} {\bibinfo {title} {{Weak Plaquette Valence Bond Order in the $S =
  1/2$ Honeycomb $J_1$–$J_2$ Heisenberg Model}},\ }\href
  {https://doi.org/10.1103/PhysRevLett.110.127205} {\bibfield  {journal}
  {\bibinfo  {journal} {Phys. Rev. Lett.}\ }\textbf {\bibinfo {volume} {110}},\
  \bibinfo {pages} {127205} (\bibinfo {year} {2013})}\BibitemShut {NoStop}%
\bibitem [{\citenamefont {Gong}\ \emph {et~al.}(2013)\citenamefont {Gong},
  \citenamefont {Sheng}, \citenamefont {Motrunich},\ and\ \citenamefont
  {Fisher}}]{Gong2013}%
  \BibitemOpen
  \bibfield  {author} {\bibinfo {author} {\bibfnamefont {S.-S.}\ \bibnamefont
  {Gong}}, \bibinfo {author} {\bibfnamefont {D.~N.}\ \bibnamefont {Sheng}},
  \bibinfo {author} {\bibfnamefont {O.~I.}\ \bibnamefont {Motrunich}},\ and\
  \bibinfo {author} {\bibfnamefont {M.~P.~A.}\ \bibnamefont {Fisher}},\
  }\bibfield  {title} {\bibinfo {title} {{Phase diagram of the spin-1/2
  $J_1$–$J_2$ Heisenberg model on a honeycomb lattice}},\ }\href
  {https://doi.org/10.1103/PhysRevB.88.165138} {\bibfield  {journal} {\bibinfo
  {journal} {Phys. Rev. B}\ }\textbf {\bibinfo {volume} {88}},\ \bibinfo
  {pages} {165138} (\bibinfo {year} {2013})}\BibitemShut {NoStop}%
\bibitem [{\citenamefont {Ghorbani}\ \emph {et~al.}(2016)\citenamefont
  {Ghorbani}, \citenamefont {Shahbazi},\ and\ \citenamefont
  {Mosadeq}}]{Ghorbani2016}%
  \BibitemOpen
  \bibfield  {author} {\bibinfo {author} {\bibfnamefont {E.}~\bibnamefont
  {Ghorbani}}, \bibinfo {author} {\bibfnamefont {F.}~\bibnamefont {Shahbazi}},\
  and\ \bibinfo {author} {\bibfnamefont {H.}~\bibnamefont {Mosadeq}},\
  }\bibfield  {title} {\bibinfo {title} {{Quantum phase diagram of distorted
  $J_1$–$J_2$ Heisenberg $S=1/2$ antiferromagnet in honeycomb lattice: a
  modified spin wave study}},\ }\href
  {https://doi.org/10.1088/0953-8984/28/40/406001} {\bibfield  {journal}
  {\bibinfo  {journal} {J. Phys.: Condens. Matter}\ }\textbf {\bibinfo {volume}
  {28}},\ \bibinfo {pages} {406001} (\bibinfo {year} {2016})}\BibitemShut
  {NoStop}%
\bibitem [{\citenamefont {Pershoguba}\ \emph {et~al.}(2018)\citenamefont
  {Pershoguba}, \citenamefont {Banerjee}, \citenamefont {Lashley},
  \citenamefont {Park}, \citenamefont {Ågren}, \citenamefont {Aeppli},\ and\
  \citenamefont {Balatsky}}]{Pershoguba2018}%
  \BibitemOpen
  \bibfield  {author} {\bibinfo {author} {\bibfnamefont {S.~S.}\ \bibnamefont
  {Pershoguba}}, \bibinfo {author} {\bibfnamefont {S.}~\bibnamefont
  {Banerjee}}, \bibinfo {author} {\bibfnamefont {J.~C.}\ \bibnamefont
  {Lashley}}, \bibinfo {author} {\bibfnamefont {J.}~\bibnamefont {Park}},
  \bibinfo {author} {\bibfnamefont {H.}~\bibnamefont {Ågren}}, \bibinfo
  {author} {\bibfnamefont {G.}~\bibnamefont {Aeppli}},\ and\ \bibinfo {author}
  {\bibfnamefont {A.~V.}\ \bibnamefont {Balatsky}},\ }\bibfield  {title}
  {\bibinfo {title} {{Dirac Magnons in Honeycomb Ferromagnets}},\ }\href
  {https://doi.org/10.1103/PhysRevX.8.011010} {\bibfield  {journal} {\bibinfo
  {journal} {Phys. Rev. X}\ }\textbf {\bibinfo {volume} {8}},\ \bibinfo {pages}
  {011010} (\bibinfo {year} {2018})}\BibitemShut {NoStop}%
\bibitem [{\citenamefont {McClarty}(2022)}]{McClarty2022}%
  \BibitemOpen
  \bibfield  {author} {\bibinfo {author} {\bibfnamefont {P.}~\bibnamefont
  {McClarty}},\ }\bibfield  {title} {\bibinfo {title} {{Topological magnons: A
  review}},\ }\href {https://doi.org/10.1146/annurev-conmatphys-031620-104715}
  {\bibfield  {journal} {\bibinfo  {journal} {Annu. Rev. Condens. Matter
  Phys.}\ }\textbf {\bibinfo {volume} {13}},\ \bibinfo {pages} {171} (\bibinfo
  {year} {2022})}\BibitemShut {NoStop}%
\bibitem [{\citenamefont {Nikitin}\ \emph {et~al.}(2022)\citenamefont
  {Nikitin}, \citenamefont {Fåk}, \citenamefont {Krämer}, \citenamefont
  {Fennell}, \citenamefont {Normand}, \citenamefont {Läuchli},\ and\
  \citenamefont {Rüegg}}]{Nikitin2022}%
  \BibitemOpen
  \bibfield  {author} {\bibinfo {author} {\bibfnamefont {S.~E.}\ \bibnamefont
  {Nikitin}}, \bibinfo {author} {\bibfnamefont {B.}~\bibnamefont {Fåk}},
  \bibinfo {author} {\bibfnamefont {K.~W.}\ \bibnamefont {Krämer}}, \bibinfo
  {author} {\bibfnamefont {T.}~\bibnamefont {Fennell}}, \bibinfo {author}
  {\bibfnamefont {B.}~\bibnamefont {Normand}}, \bibinfo {author} {\bibfnamefont
  {A.~M.}\ \bibnamefont {Läuchli}},\ and\ \bibinfo {author} {\bibfnamefont
  {C.}~\bibnamefont {Rüegg}},\ }\bibfield  {title} {\bibinfo {title} {{Thermal
  Evolution of Dirac Magnons in the Honeycomb Ferromagnet CrBr$_3$}},\ }\href
  {https://doi.org/10.1103/PhysRevLett.129.127201} {\bibfield  {journal}
  {\bibinfo  {journal} {Phys. Rev. Lett.}\ }\textbf {\bibinfo {volume} {129}},\
  \bibinfo {pages} {127201} (\bibinfo {year} {2022})}\BibitemShut {NoStop}%
\bibitem [{\citenamefont {Maksimov}\ and\ \citenamefont
  {Chernyshev}(2016)}]{mc2016}%
  \BibitemOpen
  \bibfield  {author} {\bibinfo {author} {\bibfnamefont {P.~A.}\ \bibnamefont
  {Maksimov}}\ and\ \bibinfo {author} {\bibfnamefont {A.~L.}\ \bibnamefont
  {Chernyshev}},\ }\bibfield  {title} {\bibinfo {title} {{Field-induced
  dynamical properties of the XXZ model on a honeycomb lattice}},\ }\href
  {https://doi.org/10.1103/PhysRevB.93.014418} {\bibfield  {journal} {\bibinfo
  {journal} {Phys. Rev. B}\ }\textbf {\bibinfo {volume} {93}},\ \bibinfo
  {pages} {014418} (\bibinfo {year} {2016})}\BibitemShut {NoStop}%
\bibitem [{\citenamefont {Ferrari}\ and\ \citenamefont
  {Becca}(2020)}]{Ferrari2020}%
  \BibitemOpen
  \bibfield  {author} {\bibinfo {author} {\bibfnamefont {F.}~\bibnamefont
  {Ferrari}}\ and\ \bibinfo {author} {\bibfnamefont {F.}~\bibnamefont
  {Becca}},\ }\bibfield  {title} {\bibinfo {title} {{Dynamical properties of
  N\'eel and valence-bond phases in the $J_1$–$J_2$ model on the honeycomb
  lattice}},\ }\href {https://doi.org/10.1088/1361-648X/ab7f6e} {\bibfield
  {journal} {\bibinfo  {journal} {J. Phys.: Condens. Matter}\ }\textbf
  {\bibinfo {volume} {32}},\ \bibinfo {pages} {274003} (\bibinfo {year}
  {2020})}\BibitemShut {NoStop}%
\bibitem [{\citenamefont {Gu}\ \emph {et~al.}(2022)\citenamefont {Gu},
  \citenamefont {Yu},\ and\ \citenamefont {Li}}]{Gu2022}%
  \BibitemOpen
  \bibfield  {author} {\bibinfo {author} {\bibfnamefont {C.}~\bibnamefont
  {Gu}}, \bibinfo {author} {\bibfnamefont {S.-L.}\ \bibnamefont {Yu}},\ and\
  \bibinfo {author} {\bibfnamefont {J.-X.}\ \bibnamefont {Li}},\ }\bibfield
  {title} {\bibinfo {title} {{Spin dynamics and continuum spectra of the
  honeycomb $J_1$-$J_2$ antiferromagnetic Heisenberg model}},\ }\href
  {https://doi.org/10.1103/PhysRevB.105.174403} {\bibfield  {journal} {\bibinfo
   {journal} {Phys. Rev. B}\ }\textbf {\bibinfo {volume} {105}},\ \bibinfo
  {pages} {174403} (\bibinfo {year} {2022})}\BibitemShut {NoStop}%
\bibitem [{\citenamefont {Wessler}\ \emph {et~al.}(2020)\citenamefont
  {Wessler}, \citenamefont {Roessli}, \citenamefont {Krämer}, \citenamefont
  {Delley}, \citenamefont {Waldmann}, \citenamefont {Keller}, \citenamefont
  {Cheptiakov}, \citenamefont {Braun},\ and\ \citenamefont
  {Kenzelmann}}]{Wessler2020}%
  \BibitemOpen
  \bibfield  {author} {\bibinfo {author} {\bibfnamefont {C.}~\bibnamefont
  {Wessler}}, \bibinfo {author} {\bibfnamefont {B.}~\bibnamefont {Roessli}},
  \bibinfo {author} {\bibfnamefont {K.~W.}\ \bibnamefont {Krämer}}, \bibinfo
  {author} {\bibfnamefont {B.}~\bibnamefont {Delley}}, \bibinfo {author}
  {\bibfnamefont {O.}~\bibnamefont {Waldmann}}, \bibinfo {author}
  {\bibfnamefont {L.}~\bibnamefont {Keller}}, \bibinfo {author} {\bibfnamefont
  {D.}~\bibnamefont {Cheptiakov}}, \bibinfo {author} {\bibfnamefont {H.~B.}\
  \bibnamefont {Braun}},\ and\ \bibinfo {author} {\bibfnamefont
  {M.}~\bibnamefont {Kenzelmann}},\ }\bibfield  {title} {\bibinfo {title}
  {Observation of plaquette fluctuations in the spin-1/2 honeycomb lattice},\
  }\href {https://doi.org/https://doi.org/10.1038/s41535-020-00287-1}
  {\bibfield  {journal} {\bibinfo  {journal} {npj Quantum Mater.}\ }\textbf
  {\bibinfo {volume} {5}},\ \bibinfo {pages} {85} (\bibinfo {year}
  {2020})}\BibitemShut {NoStop}%
\bibitem [{\citenamefont {Sala}\ \emph {et~al.}(2021)\citenamefont {Sala},
  \citenamefont {Stone}, \citenamefont {Rai}, \citenamefont {May},
  \citenamefont {Laurell}, \citenamefont {Garlea}, \citenamefont {Butch},
  \citenamefont {Lumsden}, \citenamefont {Ehlers}, \citenamefont {Pokharel},
  \citenamefont {Podlesnyak}, \citenamefont {Mandrus}, \citenamefont {Parker},
  \citenamefont {Okamoto}, \citenamefont {Halász},\ and\ \citenamefont
  {Christianson}}]{Sala2021}%
  \BibitemOpen
  \bibfield  {author} {\bibinfo {author} {\bibfnamefont {G.}~\bibnamefont
  {Sala}}, \bibinfo {author} {\bibfnamefont {M.~B.}\ \bibnamefont {Stone}},
  \bibinfo {author} {\bibfnamefont {B.~K.}\ \bibnamefont {Rai}}, \bibinfo
  {author} {\bibfnamefont {A.~F.}\ \bibnamefont {May}}, \bibinfo {author}
  {\bibfnamefont {P.}~\bibnamefont {Laurell}}, \bibinfo {author} {\bibfnamefont
  {V.~O.}\ \bibnamefont {Garlea}}, \bibinfo {author} {\bibfnamefont {N.~P.}\
  \bibnamefont {Butch}}, \bibinfo {author} {\bibfnamefont {M.~D.}\ \bibnamefont
  {Lumsden}}, \bibinfo {author} {\bibfnamefont {G.}~\bibnamefont {Ehlers}},
  \bibinfo {author} {\bibfnamefont {G.}~\bibnamefont {Pokharel}}, \bibinfo
  {author} {\bibfnamefont {A.}~\bibnamefont {Podlesnyak}}, \bibinfo {author}
  {\bibfnamefont {D.}~\bibnamefont {Mandrus}}, \bibinfo {author} {\bibfnamefont
  {D.~S.}\ \bibnamefont {Parker}}, \bibinfo {author} {\bibfnamefont
  {S.}~\bibnamefont {Okamoto}}, \bibinfo {author} {\bibfnamefont {G.~B.}\
  \bibnamefont {Halász}},\ and\ \bibinfo {author} {\bibfnamefont {A.~D.}\
  \bibnamefont {Christianson}},\ }\bibfield  {title} {\bibinfo {title} {{Van
  Hove singularity in the magnon spectrum of the antiferromagnetic quantum
  honeycomb lattice}},\ }\href {https://doi.org/10.1038/s41467-020-20335-5}
  {\bibfield  {journal} {\bibinfo  {journal} {Nat. Commun.}\ }\textbf {\bibinfo
  {volume} {12}},\ \bibinfo {pages} {171} (\bibinfo {year} {2021})}\BibitemShut
  {NoStop}%
\bibitem [{\citenamefont {Sala}\ \emph {et~al.}(2023)\citenamefont {Sala},
  \citenamefont {Stone}, \citenamefont {Halász}, \citenamefont {Lumsden},
  \citenamefont {May}, \citenamefont {Pajerowski}, \citenamefont
  {Ohira-Kawamura}, \citenamefont {Kaneko}, \citenamefont {Mazzone},
  \citenamefont {Simutis}, \citenamefont {Lass}, \citenamefont {Kato},
  \citenamefont {Do}, \citenamefont {Lin},\ and\ \citenamefont
  {Christianson}}]{Sala2023}%
  \BibitemOpen
  \bibfield  {author} {\bibinfo {author} {\bibfnamefont {G.}~\bibnamefont
  {Sala}}, \bibinfo {author} {\bibfnamefont {M.~B.}\ \bibnamefont {Stone}},
  \bibinfo {author} {\bibfnamefont {G.~B.}\ \bibnamefont {Halász}}, \bibinfo
  {author} {\bibfnamefont {M.~D.}\ \bibnamefont {Lumsden}}, \bibinfo {author}
  {\bibfnamefont {A.~F.}\ \bibnamefont {May}}, \bibinfo {author} {\bibfnamefont
  {D.~M.}\ \bibnamefont {Pajerowski}}, \bibinfo {author} {\bibfnamefont
  {S.}~\bibnamefont {Ohira-Kawamura}}, \bibinfo {author} {\bibfnamefont
  {K.}~\bibnamefont {Kaneko}}, \bibinfo {author} {\bibfnamefont {D.~G.}\
  \bibnamefont {Mazzone}}, \bibinfo {author} {\bibfnamefont {G.}~\bibnamefont
  {Simutis}}, \bibinfo {author} {\bibfnamefont {J.}~\bibnamefont {Lass}},
  \bibinfo {author} {\bibfnamefont {Y.}~\bibnamefont {Kato}}, \bibinfo {author}
  {\bibfnamefont {S.-H.}\ \bibnamefont {Do}}, \bibinfo {author} {\bibfnamefont
  {J.~Y.~Y.}\ \bibnamefont {Lin}},\ and\ \bibinfo {author} {\bibfnamefont
  {A.~D.}\ \bibnamefont {Christianson}},\ }\bibfield  {title} {\bibinfo {title}
  {{Field-tuned quantum renormalization of spin dynamics in the honeycomb
  lattice Heisenberg antiferromagnet YbCl$_3$}},\ }\href
  {https://doi.org/10.1038/s42005-023-01333-7} {\bibfield  {journal} {\bibinfo
  {journal} {Comms. Phys.}\ }\textbf {\bibinfo {volume} {6}},\ \bibinfo {pages}
  {234} (\bibinfo {year} {2023})}\BibitemShut {NoStop}%
\bibitem [{sm()}]{sm}%
  \BibitemOpen
  \href@noop {} {}\bibinfo {note} {The Supplemental Material at [URL inserted
  by publisher], which includes Refs.~\cite{Kraemer1999, BiI3a, BiI3b, AlCl3,
  Collins1989, Lass2020, Facheris2022, Lass2024, moon1969, Boothroyd2020,
  Wildes2006, sunny, Wessler2022, Schollwoeck2011, Stoudenmire2012,
  Hauschild2018, Popovici1975, Eckold2014}, contains detailed descriptions of
  our single-crystal sample, our unpolarized and polarized INS experiments,
  data-analysis procedures, spin-wave fits to the fully polarized spectra, and
  cylinder MPS calculations}\BibitemShut {NoStop}%
\bibitem [{\citenamefont {Lass}\ \emph {et~al.}(2023)\citenamefont {Lass},
  \citenamefont {Jacobsen}, \citenamefont {Krighaar}, \citenamefont {Graf},
  \citenamefont {Groitl}, \citenamefont {Herzog}, \citenamefont {Yamada},
  \citenamefont {K\"{a}gi}, \citenamefont {M\"{u}ller}, \citenamefont
  {B\"{u}rge}, \citenamefont {Schild}, \citenamefont {Lehmann}, \citenamefont
  {Bollhalder}, \citenamefont {Keller}, \citenamefont {Bartkowiak},
  \citenamefont {Filges}, \citenamefont {Greuter}, \citenamefont {Theidel},
  \citenamefont {R{\o}nnow}, \citenamefont {Niedermayer},\ and\ \citenamefont
  {Mazzone}}]{Lass2023}%
  \BibitemOpen
  \bibfield  {author} {\bibinfo {author} {\bibfnamefont {J.}~\bibnamefont
  {Lass}}, \bibinfo {author} {\bibfnamefont {H.}~\bibnamefont {Jacobsen}},
  \bibinfo {author} {\bibfnamefont {K.~M.~L.}\ \bibnamefont {Krighaar}},
  \bibinfo {author} {\bibfnamefont {D.}~\bibnamefont {Graf}}, \bibinfo {author}
  {\bibfnamefont {F.}~\bibnamefont {Groitl}}, \bibinfo {author} {\bibfnamefont
  {F.}~\bibnamefont {Herzog}}, \bibinfo {author} {\bibfnamefont
  {M.}~\bibnamefont {Yamada}}, \bibinfo {author} {\bibfnamefont
  {C.}~\bibnamefont {K\"{a}gi}}, \bibinfo {author} {\bibfnamefont {R.~A.}\
  \bibnamefont {M\"{u}ller}}, \bibinfo {author} {\bibfnamefont
  {R.}~\bibnamefont {B\"{u}rge}}, \bibinfo {author} {\bibfnamefont
  {M.}~\bibnamefont {Schild}}, \bibinfo {author} {\bibfnamefont {M.~S.}\
  \bibnamefont {Lehmann}}, \bibinfo {author} {\bibfnamefont {A.}~\bibnamefont
  {Bollhalder}}, \bibinfo {author} {\bibfnamefont {P.}~\bibnamefont {Keller}},
  \bibinfo {author} {\bibfnamefont {M.}~\bibnamefont {Bartkowiak}}, \bibinfo
  {author} {\bibfnamefont {U.}~\bibnamefont {Filges}}, \bibinfo {author}
  {\bibfnamefont {U.}~\bibnamefont {Greuter}}, \bibinfo {author} {\bibfnamefont
  {G.}~\bibnamefont {Theidel}}, \bibinfo {author} {\bibfnamefont {H.~M.}\
  \bibnamefont {R{\o}nnow}}, \bibinfo {author} {\bibfnamefont {C.}~\bibnamefont
  {Niedermayer}},\ and\ \bibinfo {author} {\bibfnamefont {D.~G.}\ \bibnamefont
  {Mazzone}},\ }\bibfield  {title} {\bibinfo {title} {{Commissioning of the
  novel Continuous Angle Multi-energy Analysis spectrometer at the Paul
  Scherrer Institut}},\ }\href {https://doi.org/10.1063/5.0128226} {\bibfield
  {journal} {\bibinfo  {journal} {Rev. Sci. Instrum.}\ }\textbf {\bibinfo
  {volume} {94}},\ \bibinfo {pages} {023302} (\bibinfo {year}
  {2023})}\BibitemShut {NoStop}%
\bibitem [{\citenamefont {Schmalzl}\ \emph {et~al.}(2016)\citenamefont
  {Schmalzl}, \citenamefont {Schmidt}, \citenamefont {Raymond}, \citenamefont
  {Feilbach}, \citenamefont {Mounier}, \citenamefont {Vettard},\ and\
  \citenamefont {Brückel}}]{Schmalzl2016}%
  \BibitemOpen
  \bibfield  {author} {\bibinfo {author} {\bibfnamefont {K.}~\bibnamefont
  {Schmalzl}}, \bibinfo {author} {\bibfnamefont {W.}~\bibnamefont {Schmidt}},
  \bibinfo {author} {\bibfnamefont {S.}~\bibnamefont {Raymond}}, \bibinfo
  {author} {\bibfnamefont {H.}~\bibnamefont {Feilbach}}, \bibinfo {author}
  {\bibfnamefont {C.}~\bibnamefont {Mounier}}, \bibinfo {author} {\bibfnamefont
  {B.}~\bibnamefont {Vettard}},\ and\ \bibinfo {author} {\bibfnamefont
  {T.}~\bibnamefont {Brückel}},\ }\bibfield  {title} {\bibinfo {title} {{The
  upgrade of the cold neutron three-axis spectrometer IN12 at the ILL}},\
  }\href {https://doi.org/https://doi.org/10.1016/j.nima.2016.02.067}
  {\bibfield  {journal} {\bibinfo  {journal} {Nucl. Instrum. Meth. Phys. Res.
  A}\ }\textbf {\bibinfo {volume} {819}},\ \bibinfo {pages} {89} (\bibinfo
  {year} {2016})}\BibitemShut {NoStop}%
\bibitem [{\citenamefont {Gohlke}\ \emph {et~al.}(2017)\citenamefont {Gohlke},
  \citenamefont {Verresen}, \citenamefont {Moessner},\ and\ \citenamefont
  {Pollmann}}]{Gohlke2017}%
  \BibitemOpen
  \bibfield  {author} {\bibinfo {author} {\bibfnamefont {M.}~\bibnamefont
  {Gohlke}}, \bibinfo {author} {\bibfnamefont {R.}~\bibnamefont {Verresen}},
  \bibinfo {author} {\bibfnamefont {R.}~\bibnamefont {Moessner}},\ and\
  \bibinfo {author} {\bibfnamefont {F.}~\bibnamefont {Pollmann}},\ }\bibfield
  {title} {\bibinfo {title} {{Dynamics of the Kitaev-Heisenberg Model}},\
  }\href {https://doi.org/10.1103/PhysRevLett.119.157203} {\bibfield  {journal}
  {\bibinfo  {journal} {Phys. Rev. Lett.}\ }\textbf {\bibinfo {volume} {119}},\
  \bibinfo {pages} {157203} (\bibinfo {year} {2017})}\BibitemShut {NoStop}%
\bibitem [{\citenamefont {Verresen}\ \emph {et~al.}(2018)\citenamefont
  {Verresen}, \citenamefont {Pollmann},\ and\ \citenamefont
  {Moessner}}]{Verresen2018}%
  \BibitemOpen
  \bibfield  {author} {\bibinfo {author} {\bibfnamefont {R.}~\bibnamefont
  {Verresen}}, \bibinfo {author} {\bibfnamefont {F.}~\bibnamefont {Pollmann}},\
  and\ \bibinfo {author} {\bibfnamefont {R.}~\bibnamefont {Moessner}},\
  }\bibfield  {title} {\bibinfo {title} {{Quantum dynamics of the
  square-lattice Heisenberg model}},\ }\href
  {https://doi.org/10.1103/PhysRevB.98.155102} {\bibfield  {journal} {\bibinfo
  {journal} {Phys. Rev. B}\ }\textbf {\bibinfo {volume} {98}},\ \bibinfo
  {pages} {155102} (\bibinfo {year} {2018})}\BibitemShut {NoStop}%
\bibitem [{\citenamefont {Xie}\ \emph {et~al.}(2023)\citenamefont {Xie},
  \citenamefont {Eberharter}, \citenamefont {Xing}, \citenamefont {Nishimoto},
  \citenamefont {Brando}, \citenamefont {Khanenko}, \citenamefont
  {Sichelschmidt}, \citenamefont {Turrini}, \citenamefont {Mazzone},
  \citenamefont {Naumov}, \citenamefont {Sanjeewa}, \citenamefont {Harrison},
  \citenamefont {Sefat}, \citenamefont {Normand}, \citenamefont {Läuchli},
  \citenamefont {Podlesnyak},\ and\ \citenamefont {Nikitin}}]{Xie2023}%
  \BibitemOpen
  \bibfield  {author} {\bibinfo {author} {\bibfnamefont {T.}~\bibnamefont
  {Xie}}, \bibinfo {author} {\bibfnamefont {A.~A.}\ \bibnamefont {Eberharter}},
  \bibinfo {author} {\bibfnamefont {J.}~\bibnamefont {Xing}}, \bibinfo {author}
  {\bibfnamefont {S.}~\bibnamefont {Nishimoto}}, \bibinfo {author}
  {\bibfnamefont {M.}~\bibnamefont {Brando}}, \bibinfo {author} {\bibfnamefont
  {P.}~\bibnamefont {Khanenko}}, \bibinfo {author} {\bibfnamefont
  {J.}~\bibnamefont {Sichelschmidt}}, \bibinfo {author} {\bibfnamefont {A.~A.}\
  \bibnamefont {Turrini}}, \bibinfo {author} {\bibfnamefont {D.~G.}\
  \bibnamefont {Mazzone}}, \bibinfo {author} {\bibfnamefont {P.~G.}\
  \bibnamefont {Naumov}}, \bibinfo {author} {\bibfnamefont {L.~D.}\
  \bibnamefont {Sanjeewa}}, \bibinfo {author} {\bibfnamefont {N.}~\bibnamefont
  {Harrison}}, \bibinfo {author} {\bibfnamefont {A.~S.}\ \bibnamefont {Sefat}},
  \bibinfo {author} {\bibfnamefont {B.}~\bibnamefont {Normand}}, \bibinfo
  {author} {\bibfnamefont {A.~M.}\ \bibnamefont {Läuchli}}, \bibinfo {author}
  {\bibfnamefont {A.}~\bibnamefont {Podlesnyak}},\ and\ \bibinfo {author}
  {\bibfnamefont {S.~E.}\ \bibnamefont {Nikitin}},\ }\bibfield  {title}
  {\bibinfo {title} {{Complete field-induced spectral response of the spin-1/2
  triangular-lattice antiferromagnet CsYbSe$_2$}},\ }\href
  {https://doi.org/10.1038/s41535-023-00580-9} {\bibfield  {journal} {\bibinfo
  {journal} {npj Quantum Mater.}\ }\textbf {\bibinfo {volume} {8}},\ \bibinfo
  {pages} {48} (\bibinfo {year} {2023})}\BibitemShut {NoStop}%
\bibitem [{\citenamefont {Weber}\ \emph {et~al.}(2016)\citenamefont {Weber},
  \citenamefont {Georgii},\ and\ \citenamefont {Böni}}]{Weber2016}%
  \BibitemOpen
  \bibfield  {author} {\bibinfo {author} {\bibfnamefont {T.}~\bibnamefont
  {Weber}}, \bibinfo {author} {\bibfnamefont {R.}~\bibnamefont {Georgii}},\
  and\ \bibinfo {author} {\bibfnamefont {P.}~\bibnamefont {Böni}},\ }\bibfield
   {title} {\bibinfo {title} {{Takin: An open-source software for experiment
  planning, visualisation, and data analysis}},\ }\href
  {https://doi.org/https://doi.org/10.1016/j.softx.2016.06.002} {\bibfield
  {journal} {\bibinfo  {journal} {SoftwareX}\ }\textbf {\bibinfo {volume}
  {5}},\ \bibinfo {pages} {121} (\bibinfo {year} {2016})}\BibitemShut {NoStop}%
\bibitem [{\citenamefont {Pocs}\ \emph {et~al.}(2021)\citenamefont {Pocs},
  \citenamefont {Siegfried}, \citenamefont {Xing}, \citenamefont {Sefat},
  \citenamefont {Hermele}, \citenamefont {Normand},\ and\ \citenamefont
  {Lee}}]{Pocs2021}%
  \BibitemOpen
  \bibfield  {author} {\bibinfo {author} {\bibfnamefont {C.~A.}\ \bibnamefont
  {Pocs}}, \bibinfo {author} {\bibfnamefont {P.~E.}\ \bibnamefont {Siegfried}},
  \bibinfo {author} {\bibfnamefont {J.}~\bibnamefont {Xing}}, \bibinfo {author}
  {\bibfnamefont {A.~S.}\ \bibnamefont {Sefat}}, \bibinfo {author}
  {\bibfnamefont {M.}~\bibnamefont {Hermele}}, \bibinfo {author} {\bibfnamefont
  {B.}~\bibnamefont {Normand}},\ and\ \bibinfo {author} {\bibfnamefont
  {M.}~\bibnamefont {Lee}},\ }\bibfield  {title} {\bibinfo {title} {{Systematic
  extraction of crystal electric-field effects and quantum magnetic model
  parameters in triangular rare-earth magnets}},\ }\href
  {https://doi.org/10.1103/PhysRevResearch.3.043202} {\bibfield  {journal}
  {\bibinfo  {journal} {Phys. Rev. Research}\ }\textbf {\bibinfo {volume}
  {3}},\ \bibinfo {pages} {043202} (\bibinfo {year} {2021})}\BibitemShut
  {NoStop}%
\bibitem [{\citenamefont {Oshikawa}\ and\ \citenamefont
  {Affleck}(1997)}]{oshikawa1997}%
  \BibitemOpen
  \bibfield  {author} {\bibinfo {author} {\bibfnamefont {M.}~\bibnamefont
  {Oshikawa}}\ and\ \bibinfo {author} {\bibfnamefont {I.}~\bibnamefont
  {Affleck}},\ }\bibfield  {title} {\bibinfo {title} {{Field-Induced Gap in $S
  = 1/2$ Antiferromagnetic Chains}},\ }\href
  {https://doi.org/10.1103/PhysRevLett.79.2883} {\bibfield  {journal} {\bibinfo
   {journal} {Phys. Rev. Lett.}\ }\textbf {\bibinfo {volume} {79}},\ \bibinfo
  {pages} {2883} (\bibinfo {year} {1997})}\BibitemShut {NoStop}%
\bibitem [{\citenamefont {Rau}\ and\ \citenamefont
  {Gingras}(2018)}]{RauGingras2018}%
  \BibitemOpen
  \bibfield  {author} {\bibinfo {author} {\bibfnamefont {J.~G.}\ \bibnamefont
  {Rau}}\ and\ \bibinfo {author} {\bibfnamefont {M.~J.~P.}\ \bibnamefont
  {Gingras}},\ }\bibfield  {title} {\bibinfo {title} {Frustration and
  anisotropic exchange in ytterbium magnets with edge-shared octahedra},\
  }\href {https://doi.org/10.1103/PhysRevB.98.054408} {\bibfield  {journal}
  {\bibinfo  {journal} {Phys. Rev. B}\ }\textbf {\bibinfo {volume} {98}},\
  \bibinfo {pages} {054408} (\bibinfo {year} {2018})}\BibitemShut {NoStop}%
\bibitem [{\citenamefont {Verresen}\ \emph {et~al.}(2019)\citenamefont
  {Verresen}, \citenamefont {Moessner},\ and\ \citenamefont
  {Pollmann}}]{Verresen2019}%
  \BibitemOpen
  \bibfield  {author} {\bibinfo {author} {\bibfnamefont {R.}~\bibnamefont
  {Verresen}}, \bibinfo {author} {\bibfnamefont {R.}~\bibnamefont {Moessner}},\
  and\ \bibinfo {author} {\bibfnamefont {F.}~\bibnamefont {Pollmann}},\
  }\bibfield  {title} {\bibinfo {title} {Avoided quasiparticle decay from
  strong quantum interactions},\ }\href
  {https://doi.org/10.1038/s41567-019-0535-3} {\bibfield  {journal} {\bibinfo
  {journal} {Nat. Phys.}\ }\textbf {\bibinfo {volume} {15}},\ \bibinfo {pages}
  {750} (\bibinfo {year} {2019})}\BibitemShut {NoStop}%
\bibitem [{sls()}]{slsm2024}%
  \BibitemOpen
  \href@noop {} {}\bibinfo {note} {F. Elson, M. Nayak, A. A. Eberharter, M.
  Skoulatos, S. Ward, N. B. Christensen, D. Voneshen, C. Fiolka, K. W. Krämer,
  Ch. Rüegg, B. Normand, M. Mourigal, H. M. Rønnow, F. Mila, A. M. Läuchli
  and M. Månsson, Field-induced shadow-mode excitations in the square-lattice
  quantum antiferromagnet CuF$_2$(D$_2$O)$_2$(pyz), unpublished}\BibitemShut
  {NoStop}%
\bibitem [{\citenamefont {Landau}(1941)}]{Landau1941}%
  \BibitemOpen
  \bibfield  {author} {\bibinfo {author} {\bibfnamefont {L.~D.}\ \bibnamefont
  {Landau}},\ }\bibfield  {title} {\bibinfo {title} {{Theory of the
  Superfluidity of Helium II}},\ }\href
  {https://doi.org/10.1103/PhysRev.60.356} {\bibfield  {journal} {\bibinfo
  {journal} {Phys. Rev.}\ }\textbf {\bibinfo {volume} {60}},\ \bibinfo {pages}
  {356} (\bibinfo {year} {1941})}\BibitemShut {NoStop}%
\bibitem [{ill()}]{ill}%
  \BibitemOpen
  \bibinfo {title} {{J. A. Hernández, A. A. Eberharter, O. S. Fjellvag, M.
  Kenzelmann, K. W. Krämer, A. M. Läuchli, S. Raymond, B. Roessli, and M.
  Schuler, Investigation of the Polarization-Dependence of Field-Induced Roton
  Minima in the 2D van der Waals Honeycomb Antiferromagnet YbBr$_3$, (2022),
  \url{https://doi.ill.fr/10.5291/ILL-DATA.CRG-3010}},}\BibitemShut {NoStop}%
\bibitem [{\citenamefont {Kr\"amer}\ \emph {et~al.}(1999)\citenamefont
  {Kr\"amer}, \citenamefont {G\"udel}, \citenamefont {Roessli}, \citenamefont
  {Fischer}, \citenamefont {D\"onni}, \citenamefont {Wada}, \citenamefont
  {Fauth}, \citenamefont {Fernandez-Diaz},\ and\ \citenamefont
  {Hauss}}]{Kraemer1999}%
  \BibitemOpen
\bibfield  {title} {  }\bibfield  {author} {\bibinfo {author} {\bibfnamefont
  {K.~W.}\ \bibnamefont {Kr\"amer}}, \bibinfo {author} {\bibfnamefont {H.~U.}\
  \bibnamefont {G\"udel}}, \bibinfo {author} {\bibfnamefont {B.}~\bibnamefont
  {Roessli}}, \bibinfo {author} {\bibfnamefont {P.}~\bibnamefont {Fischer}},
  \bibinfo {author} {\bibfnamefont {A.}~\bibnamefont {D\"onni}}, \bibinfo
  {author} {\bibfnamefont {N.}~\bibnamefont {Wada}}, \bibinfo {author}
  {\bibfnamefont {F.}~\bibnamefont {Fauth}}, \bibinfo {author} {\bibfnamefont
  {M.~T.}\ \bibnamefont {Fernandez-Diaz}},\ and\ \bibinfo {author}
  {\bibfnamefont {T.}~\bibnamefont {Hauss}},\ }\bibfield  {title} {\bibinfo
  {title} {{Noncollinear two- and three-dimensional magnetic ordering in the
  honeycomb lattices of $\mathrm{Er}{X}_{3}$ ($X$ = Cl, Br, I)}},\ }\href
  {https://doi.org/10.1103/PhysRevB.60.R3724} {\bibfield  {journal} {\bibinfo
  {journal} {Phys. Rev. B}\ }\textbf {\bibinfo {volume} {60}},\ \bibinfo
  {pages} {R3724} (\bibinfo {year} {1999})}\BibitemShut {NoStop}%
\bibitem [{\citenamefont {Brown}\ \emph {et~al.}(1968)\citenamefont {Brown},
  \citenamefont {Fletcher},\ and\ \citenamefont {Holah}}]{BiI3a}%
  \BibitemOpen
  \bibfield  {author} {\bibinfo {author} {\bibfnamefont {D.}~\bibnamefont
  {Brown}}, \bibinfo {author} {\bibfnamefont {S.}~\bibnamefont {Fletcher}},\
  and\ \bibinfo {author} {\bibfnamefont {D.~G.}\ \bibnamefont {Holah}},\
  }\bibfield  {title} {\bibinfo {title} {The preparation and crystallographic
  properties of certain lanthanide and actinide tribromides and tribromide
  hexahydrates},\ }\href {https://doi.org/10.1039/J19680001889} {\bibfield
  {journal} {\bibinfo  {journal} {J. Chem. Soc. A}\ }\textbf {\bibinfo {volume}
  {2}},\ \bibinfo {pages} {1889} (\bibinfo {year} {1968})}\BibitemShut
  {NoStop}%
\bibitem [{\citenamefont {Asprey}\ \emph {et~al.}(1964)\citenamefont {Asprey},
  \citenamefont {Keenan},\ and\ \citenamefont {Kruse}}]{BiI3b}%
  \BibitemOpen
  \bibfield  {author} {\bibinfo {author} {\bibfnamefont {L.~B.}\ \bibnamefont
  {Asprey}}, \bibinfo {author} {\bibfnamefont {T.~K.}\ \bibnamefont {Keenan}},\
  and\ \bibinfo {author} {\bibfnamefont {F.~H.}\ \bibnamefont {Kruse}},\
  }\bibfield  {title} {\bibinfo {title} {{Preparation and Crystal Data for
  Lanthanide and Actinide Triiodides}},\ }\href
  {https://doi.org/10.1021/ic50018a015} {\bibfield  {journal} {\bibinfo
  {journal} {Inorg. Chem.}\ }\textbf {\bibinfo {volume} {3}},\ \bibinfo {pages}
  {1137} (\bibinfo {year} {1964})}\BibitemShut {NoStop}%
\bibitem [{\citenamefont {Templeton}\ and\ \citenamefont
  {Carter}(1954)}]{AlCl3}%
  \BibitemOpen
  \bibfield  {author} {\bibinfo {author} {\bibfnamefont {D.~H.}\ \bibnamefont
  {Templeton}}\ and\ \bibinfo {author} {\bibfnamefont {G.~F.}\ \bibnamefont
  {Carter}},\ }\bibfield  {title} {\bibinfo {title} {{The Crystal Structures of
  Yttrium Trichloride and Similar Compounds}},\ }\href
  {https://doi.org/10.1021/j150521a002} {\bibfield  {journal} {\bibinfo
  {journal} {J. Phys. Chem.}\ }\textbf {\bibinfo {volume} {58}},\ \bibinfo
  {pages} {940} (\bibinfo {year} {1954})}\BibitemShut {NoStop}%
\bibitem [{\citenamefont {Collins}(1989)}]{Collins1989}%
  \BibitemOpen
  \bibfield  {author} {\bibinfo {author} {\bibfnamefont {M.~F.}\ \bibnamefont
  {Collins}},\ }\href@noop {} {\emph {\bibinfo {title} {{Magnetic Critical
  Scattering}}}}\ (\bibinfo  {publisher} {Oxford University Press},\ \bibinfo
  {year} {1989})\BibitemShut {NoStop}%
\bibitem [{\citenamefont {Lass}\ \emph {et~al.}(2020)\citenamefont {Lass},
  \citenamefont {Jacobsen}, \citenamefont {Mazzone},\ and\ \citenamefont
  {Lefmann}}]{Lass2020}%
  \BibitemOpen
  \bibfield  {author} {\bibinfo {author} {\bibfnamefont {J.}~\bibnamefont
  {Lass}}, \bibinfo {author} {\bibfnamefont {H.}~\bibnamefont {Jacobsen}},
  \bibinfo {author} {\bibfnamefont {D.~G.}\ \bibnamefont {Mazzone}},\ and\
  \bibinfo {author} {\bibfnamefont {K.}~\bibnamefont {Lefmann}},\ }\bibfield
  {title} {\bibinfo {title} {{MJOLNIR: A software package for multiplexing
  neutron spectrometers}},\ }\href
  {https://doi.org/https://doi.org/10.1016/j.softx.2020.100600} {\bibfield
  {journal} {\bibinfo  {journal} {SoftwareX}\ }\textbf {\bibinfo {volume}
  {12}},\ \bibinfo {pages} {100600} (\bibinfo {year} {2020})}\BibitemShut
  {NoStop}%
\bibitem [{\citenamefont {Facheris}\ \emph {et~al.}(2022)\citenamefont
  {Facheris}, \citenamefont {Povarov}, \citenamefont {Nabi}, \citenamefont
  {Mazzone}, \citenamefont {Lass}, \citenamefont {Roessli}, \citenamefont
  {Ressouche}, \citenamefont {Yan}, \citenamefont {Gvasaliya},\ and\
  \citenamefont {Zheludev}}]{Facheris2022}%
  \BibitemOpen
  \bibfield  {author} {\bibinfo {author} {\bibfnamefont {L.}~\bibnamefont
  {Facheris}}, \bibinfo {author} {\bibfnamefont {K.~Y.}\ \bibnamefont
  {Povarov}}, \bibinfo {author} {\bibfnamefont {S.~D.}\ \bibnamefont {Nabi}},
  \bibinfo {author} {\bibfnamefont {D.~G.}\ \bibnamefont {Mazzone}}, \bibinfo
  {author} {\bibfnamefont {J.}~\bibnamefont {Lass}}, \bibinfo {author}
  {\bibfnamefont {B.}~\bibnamefont {Roessli}}, \bibinfo {author} {\bibfnamefont
  {E.}~\bibnamefont {Ressouche}}, \bibinfo {author} {\bibfnamefont
  {Z.}~\bibnamefont {Yan}}, \bibinfo {author} {\bibfnamefont {S.}~\bibnamefont
  {Gvasaliya}},\ and\ \bibinfo {author} {\bibfnamefont {A.}~\bibnamefont
  {Zheludev}},\ }\bibfield  {title} {\bibinfo {title} {{Spin Density Wave
  versus Fractional Magnetization Plateau in a Triangular Antiferromagnet}},\
  }\href {https://doi.org/10.1103/PhysRevLett.129.087201} {\bibfield  {journal}
  {\bibinfo  {journal} {Phys. Rev. Lett.}\ }\textbf {\bibinfo {volume} {129}},\
  \bibinfo {pages} {087201} (\bibinfo {year} {2022})}\BibitemShut {NoStop}%
\bibitem [{\citenamefont {Lass}\ \emph {et~al.}(2024)\citenamefont {Lass},
  \citenamefont {Lenander}, \citenamefont {Krighaar}, \citenamefont {Tošić},
  \citenamefont {Prabhakaran}, \citenamefont {Deen}, \citenamefont
  {Holm-Janas},\ and\ \citenamefont {Lefmann}}]{Lass2024}%
  \BibitemOpen
  \bibfield  {author} {\bibinfo {author} {\bibfnamefont {J.}~\bibnamefont
  {Lass}}, \bibinfo {author} {\bibfnamefont {E.~Y.}\ \bibnamefont {Lenander}},
  \bibinfo {author} {\bibfnamefont {K.~M.~L.}\ \bibnamefont {Krighaar}},
  \bibinfo {author} {\bibfnamefont {T.~N.}\ \bibnamefont {Tošić}}, \bibinfo
  {author} {\bibfnamefont {D.}~\bibnamefont {Prabhakaran}}, \bibinfo {author}
  {\bibfnamefont {P.~P.}\ \bibnamefont {Deen}}, \bibinfo {author}
  {\bibfnamefont {S.}~\bibnamefont {Holm-Janas}},\ and\ \bibinfo {author}
  {\bibfnamefont {K.}~\bibnamefont {Lefmann}},\ }\bibfield  {title} {\bibinfo
  {title} {{Characterizing the diffuse continuum excitations in the classical
  spin liquid $h$-YMnO$_3$}},\ }\href
  {https://doi.org/10.1103/PhysRevB.110.144429} {\bibfield  {journal} {\bibinfo
   {journal} {Phys. Rev. B}\ }\textbf {\bibinfo {volume} {110}},\ \bibinfo
  {pages} {144429} (\bibinfo {year} {2024})}\BibitemShut {NoStop}%
\bibitem [{\citenamefont {Moon}\ \emph {et~al.}(1969)\citenamefont {Moon},
  \citenamefont {Riste},\ and\ \citenamefont {Koehler}}]{moon1969}%
  \BibitemOpen
  \bibfield  {author} {\bibinfo {author} {\bibfnamefont {R.~M.}\ \bibnamefont
  {Moon}}, \bibinfo {author} {\bibfnamefont {T.}~\bibnamefont {Riste}},\ and\
  \bibinfo {author} {\bibfnamefont {W.~C.}\ \bibnamefont {Koehler}},\
  }\bibfield  {title} {\bibinfo {title} {{Polarization Analysis of
  Thermal-Neutron Scattering}},\ }\href
  {https://doi.org/10.1103/PhysRev.181.920} {\bibfield  {journal} {\bibinfo
  {journal} {Phys. Rev.}\ }\textbf {\bibinfo {volume} {181}},\ \bibinfo {pages}
  {920} (\bibinfo {year} {1969})}\BibitemShut {NoStop}%
\bibitem [{\citenamefont {Boothroyd}(2020)}]{Boothroyd2020}%
  \BibitemOpen
  \bibfield  {author} {\bibinfo {author} {\bibfnamefont {A.~T.}\ \bibnamefont
  {Boothroyd}},\ }\href@noop {} {\emph {\bibinfo {title} {{Principles of
  Neutron Scattering from Condensed Matter}}}}\ (\bibinfo  {publisher} {Oxford
  University Press},\ \bibinfo {year} {2020})\BibitemShut {NoStop}%
\bibitem [{\citenamefont {Wildes}(2006)}]{Wildes2006}%
  \BibitemOpen
  \bibfield  {author} {\bibinfo {author} {\bibfnamefont {A.~R.}\ \bibnamefont
  {Wildes}},\ }\bibfield  {title} {\bibinfo {title} {Scientific reviews:
  Neutron polarization analysis corrections made easy},\ }\href
  {https://doi.org/10.1080/10448630600668738} {\bibfield  {journal} {\bibinfo
  {journal} {Neutron News}\ }\textbf {\bibinfo {volume} {17}},\ \bibinfo
  {pages} {17} (\bibinfo {year} {2006})}\BibitemShut {NoStop}%
\bibitem [{Sunny.jl, version v0.7.3, software available online at
  \url{https://github.com/SunnySuite/Sunny.jl}()}]{sunny}%
  \BibitemOpen
  Sunny.jl, version v0.7.3, software available online at
  \url{https://github.com/SunnySuite/Sunny.jl},\ \href@noop {} {}\BibitemShut
  {NoStop}%
\bibitem [{\citenamefont {Wessler}(2022)}]{Wessler2022}%
  \BibitemOpen
  \bibfield  {author} {\bibinfo {author} {\bibfnamefont {C.}~\bibnamefont
  {Wessler}},\ }\emph {\bibinfo {title} {Neutron scattering study of the 2D
  dipolar magnet ErBr$_3$ and the 2D quantum spin liquid system YbBr$_3$}},\
  \href@noop {} {Ph.D. thesis},\ \bibinfo  {school} {Universität Basel}
  (\bibinfo {year} {2022})\BibitemShut {NoStop}%
\bibitem [{\citenamefont {Schollw\"ock}(2011)}]{Schollwoeck2011}%
  \BibitemOpen
  \bibfield  {author} {\bibinfo {author} {\bibfnamefont {U.}~\bibnamefont
  {Schollw\"ock}},\ }\bibfield  {title} {\bibinfo {title} {{The density-matrix
  renormalization group in the age of matrix product states}},\ }\href
  {https://doi.org/10.1016/j.aop.2010.09.012} {\bibfield  {journal} {\bibinfo
  {journal} {Ann. Phys.}\ }\textbf {\bibinfo {volume} {326}},\ \bibinfo {pages}
  {96} (\bibinfo {year} {2011})}\BibitemShut {NoStop}%
\bibitem [{\citenamefont {Stoudenmire}\ and\ \citenamefont
  {White}(2012)}]{Stoudenmire2012}%
  \BibitemOpen
  \bibfield  {author} {\bibinfo {author} {\bibfnamefont {E.~M.}\ \bibnamefont
  {Stoudenmire}}\ and\ \bibinfo {author} {\bibfnamefont {S.~R.}\ \bibnamefont
  {White}},\ }\bibfield  {title} {\bibinfo {title} {{Studying Two-Dimensional
  Systems with the Density Matrix Renormalization Group}},\ }\href
  {https://doi.org/10.1146/annurev-conmatphys-020911-125018} {\bibfield
  {journal} {\bibinfo  {journal} {Annu. Rev. Condens. Matter Phys.}\ }\textbf
  {\bibinfo {volume} {3}},\ \bibinfo {pages} {111} (\bibinfo {year}
  {2012})}\BibitemShut {NoStop}%
\bibitem [{\citenamefont {Hauschild}\ and\ \citenamefont
  {Pollmann}(2018)}]{Hauschild2018}%
  \BibitemOpen
  \bibfield  {author} {\bibinfo {author} {\bibfnamefont {J.}~\bibnamefont
  {Hauschild}}\ and\ \bibinfo {author} {\bibfnamefont {F.}~\bibnamefont
  {Pollmann}},\ }\bibfield  {title} {\bibinfo {title} {{Efficient numerical
  simulations with Tensor Networks: Tensor Network Python (TeNPy)}},\ }\href
  {https://doi.org/10.21468/SciPostPhysLectNotes.5} {\bibfield  {journal}
  {\bibinfo  {journal} {SciPost Phys. Lect. Notes}\ }\textbf {\bibinfo {volume}
  {5}},\ \bibinfo {pages} {1} (\bibinfo {year} {2018})}\BibitemShut {NoStop}%
\bibitem [{\citenamefont {Popovici}(1975)}]{Popovici1975}%
  \BibitemOpen
  \bibfield  {author} {\bibinfo {author} {\bibfnamefont {M.}~\bibnamefont
  {Popovici}},\ }\bibfield  {title} {\bibinfo {title} {{On the resolution of
  slow-neutron spectrometers. IV. The triple-axis spectrometer resolution
  function, spatial effects included}},\ }\href
  {https://doi.org/10.1107/S0567739475001088} {\bibfield  {journal} {\bibinfo
  {journal} {Acta Crystallogr. A}\ }\textbf {\bibinfo {volume} {31}},\ \bibinfo
  {pages} {507} (\bibinfo {year} {1975})}\BibitemShut {NoStop}%
\bibitem [{\citenamefont {Eckold}\ and\ \citenamefont
  {Sobolev}(2014)}]{Eckold2014}%
  \BibitemOpen
  \bibfield  {author} {\bibinfo {author} {\bibfnamefont {G.}~\bibnamefont
  {Eckold}}\ and\ \bibinfo {author} {\bibfnamefont {O.}~\bibnamefont
  {Sobolev}},\ }\bibfield  {title} {\bibinfo {title} {Analytical approach to
  the 4d-resolution function of three axes neutron spectrometers with focussing
  monochromators and analysers},\ }\href
  {https://doi.org/https://doi.org/10.1016/j.nima.2014.03.019} {\bibfield
  {journal} {\bibinfo  {journal} {Nucl. Instrum. Methods Phys. Res. A}\
  }\textbf {\bibinfo {volume} {752}},\ \bibinfo {pages} {54} (\bibinfo {year}
  {2014})}\BibitemShut {NoStop}%
\end{thebibliography}
\end{document}